\documentclass[12pt, a4paper]{article}
\usepackage{amsmath,amssymb,bm,epsfig,afterpage}
\usepackage{cite}
\usepackage{here}
\usepackage{color}
\usepackage{colortbl}
\usepackage{xcolor}
\usepackage{tabularx}
\usepackage{subcaption}
\usepackage{bbm}
\usepackage{graphicx}
\usepackage{listings}
\usepackage{longtable}
\usepackage[utf8]{inputenc}
\usepackage{cancel}
\usepackage{slashed}

\definecolor{Orange}{cmyk}{0,0.61,0.87,0}
\definecolor{JungleGreen}{cmyk}{0.99,0,0.52,0}
\definecolor{OliveGreen}{cmyk}{0.64,0,0.95,0.40}
\definecolor{Brown}{cmyk}{0,0.81,1,0.60}
\definecolor{RoyalBlue}{cmyk}{0.71,0.53,0,0.12}
\definecolor{DarkRed}{cmyk}{0.20,0.80, 0.3, 0.2}
\definecolor{DarkBlue}{cmyk}{0.80,0.20, 0.2, 0.2}
\definecolor{Gray}{cmyk}{0,0,0,0.40}
\definecolor{LightPink}{cmyk}{0.0,0.25,0,0}
\definecolor{LLightPink}{cmyk}{0.0,0.10,0,0}
\definecolor{LightBlue}{cmyk}{0.25,0,0,0}

\definecolor{LightGray}{cmyk}{0,0,0,0.05}
\definecolor{HighOrange}{cmyk}{0., 0.41, 0.50,0}

\newcommand{\vev}[1]{{\langle{#1}\rangle}}
\newcommand{\order}[1]{\mathcal{O}\left({#1}\right)}

\newcommand{{\Lcal}}{\mathcal{L}}
\newcommand{\Mcal}{\mathcal{M}}
\newcommand{\Acal}{\mathcal{A}}
\newcommand{\Hcal}{\mathcal{H}}
\newcommand{\Ocal}{\mathcal{O}}

\newcommand{\Zbbm}{\mathbbm{Z}}
\newcommand{\ZR}[1]{\mathbbm{Z}^R_{{#1}}}
\newcommand{\Zr}[1]{\mathbbm{Z}^R_{{#1}}}
\newcommand{\Zn}[1]{\mathbbm{Z}_{{#1}}}
\newcommand{\id}{\mathbf{1}}
\newcommand{\rep}[1]{\mathbf{#1}}

\newcommand{\be}{\begin{equation}}
\newcommand{\ee}{\end{equation}}

\newcommand{\abs}[1]{\left|{#1}\right|}

\newcommand{\ol}[1]{\overline{#1}}

\newcommand{\eps}{\varepsilon}
\newcommand{\la}{\lambda}
\newcommand{\ka}{\kappa}

\newcommand{\PS}{{\mathrm{PS}}}
\newcommand{\PQ}{{\mathrm{PQ}}}
\newcommand{\UPQ}{U(1)_{\mathrm{PQ}}}
\newcommand{\LQCD}{\Lambda_{\mathrm{QCD}}}

\newcommand{\Qcal}{\mathcal{Q} }

\newcommand{\ola}{\overline{\lambda}}
\newcommand{\oka}{\overline{\kappa}}
\newcommand{\Sig}{\Sigma}
\newcommand{\oSig}{\overline{\Sigma}}

\newcommand{\Hbu}{\overline{H}_u}
\newcommand{\Hbd}{\overline{H}_d}
\newcommand{\mhu}{m_{H_u}^2}
\newcommand{\mhd}{m_{H_d}^2}
\newcommand{\mbu}{\ol{m}_{H_u}^2}
\newcommand{\mbd}{\ol{m}_{H_d}^2}

\allowdisplaybreaks[3]
\newcolumntype{Y}{&gt;{\centering\arraybackslash}X}
\usepackage[colorlinks=true, linkcolor=OliveGreen, citecolor=RoyalBlue,
urlcolor=RoyalBlue]{hyperref}

\definecolor{darkgreen}{HTML}{109930}

\begin{document}

\begin{titlepage}

\begin{flushright}
{\tt
}
\end{flushright}

\vskip 1.35cm
\begin{center}

{\Large
{\bf
Qualities of axion and LSP in Pati-Salam unification
with $\mathbbm{Z}^R_{4}\times \mathbbm{Z}_{N}$ symmetry
}
}

\vskip 1.5cm

Junichiro~Kawamura$^{a,b,}$\footnote{%
\href{mailto:kawamura.14@osu.edu}{\tt kawamura.14@osu.edu}},
Stuart~Raby$^{a,}$\footnote{%
\href{mailto:raby.1@osu.edu}{\tt raby.1@osu.edu}},
\vskip 0.8cm

{\it $^a$Department of Physics, Ohio State University, Columbus, Ohio
 43210, USA}
\\[3pt]
{\it $^b$Department of Physics, Keio University, Yokohama 223-8522, Japan}
\\[3pt]

\date{\today}

\vskip 1.5cm

\begin{abstract}
In this paper we construct supersymmetric Pati-Salam (PS) models containing
the minimal supersymmetric standard model and an invisible axion.
The models include two discrete symmetries,  $\mathbb{Z}_4^R \times \mathbb{Z}_N$,
which maintain the $quality$ of the accidental Peccei-Quinn (PQ) symmetry
and thus the solution to the strong CP problem.
We require that the discrete anomaly conditions are satisfied for both
$\mathbb{Z}_4^R \times G_{\rm PS}^2$ and $ \mathbb{Z}_N \times G_{\rm PS}^2$.
The vacuum expectation value of the PQ field spontaneously breaks all the discrete symmetries.
R-parity is violated if any of the PQ field(s) has an odd charge under $\mathbb{Z}_4^R$.
We present two explicit models
which we refer to as a minimal model where R-parity violation is extremely suppressed,
and a non-minimal model where R-parity violation is significant.
In the latter model, the neutralino becomes unstable even if it is the Lightest Supersymmetric Particle (LSP),
and, in addition, there are new low-energy vector-like states.
In both examples, R-parity violation is sufficiently suppressed such that the proton is stable.
\end{abstract}
\end{center}

\end{titlepage}
\setcounter{footnote}{0}

\section{Introduction}

The Peccei-Quinn (PQ) symmetry, $\UPQ$,
provides an attractive solution to the strong CP problem~\cite{Peccei:1977hh,Peccei:1977ur}.
The $\theta$ angle in Quantum Chromo Dynamics (QCD) settles at zero dynamically due to
the potential of a pseudo Nambu-Goldstone boson of PQ symmetry breaking, i.e. the so-called axion,
generated through QCD quantum effects~\cite{Weinberg:1977ma,Wilczek:1977pj}.
Since the PQ symmetry is an anomalous global symmetry, it will be broken by quantum gravity effects.
However, the PQ breaking effects should be extremely suppressed
such that the QCD axion potential still has a minimum at $\abs{\theta} < 10^{-10}$
to be consistent with the measurements of the neutron electric dipole moment~\cite{Baker:2006ts,Afach:2015sja,Graner:2016ses}.
This problem is known as the axion quality problem~\cite{Georgi:1981pu,Dine:1986bg,Kamionkowski:1992mf,Holman:1992us,Ghigna:1992iv}.

In this paper, we propose simple models with Pati-Salam (PS) gauge symmetry
and non-anomalous discrete symmetries, $\Zr{4}\times\Zn{N}$, where $N$ is an integer.
We aim to construct models in which the PQ symmetry arises as an accidental symmetry
and its $quality$ is ensured by the discrete symmetries~\footnote{
See Refs.~\cite{Randall:1992ut,Izawa:2002qk,Choi:2003wr,Harigaya:2013vja,Lillard:2018fdt,Hook:2019qoh,Ardu:2020qmo,Yin:2020dfn,DiLuzio:2020qio} for solutions to the quality problem.
}.
The Pati-Salam (PS) unification~\cite{Pati:1974yy} of the Standard Model (SM) is attractive
because the SM fermions are unified into two multiplets, hypercharge is quantized, and the proton is not destabilized by exotic gauge/Higgs bosons.
Although the PS gauge symmetry, $G_\PS := SU(4)_C\times SU(2)_L\times SU(2)_R$,
is not grand unified to a simple group,
the PS model can be realized in an orbifold Grand Unification Theory (GUT) in extra dimensions and from the heterotic string, for example, see Ref. ~\cite{Kobayashi:2004ud,Kobayashi:2004ya}.   Thus gauge coupling unification can be assumed with small threshold corrections at the GUT/compactification scale.
It has been shown that the recent experimental data can be explained very precisely
in the PS model~\cite{Poh:2015wta,Poh:2017xvg}.

We will consider PS models with supersymmetry (SUSY) and the discrete R-symmetry.
The Minimal Supersymmetric Standard Model (MSSM) is an attractive candidate
for a model at the TeV scale, since it solves the gauge hierarchy problem,
three gauge couplings constants are unified at the GUT scale
and the electroweak (EW) symmetry breaking is triggered radiatively.
The $\Zr{4}$ symmetry is a unique anomaly-free symmetry consistent with the PS unification
which can forbid the dimension-4 and dimension-5 operators responsible for proton decay,
as well as the mass term of the Higgs doublets at the Planck scale~\cite{Lee:2010gv,Lee:2011dya}.
Without the PQ field, R-parity exists exactly if the $\Zr{4}$ symmetry is broken by non-perturbative effects
associated with SUSY breaking, and thus the Lightest Supersymmetric Particle (LSP) will contribute to the Dark Matter (DM).

We also introduce a non-anomalous $\Zn{N}$ symmetry
to solve the axion quality problem.
Since the PQ field carries charges under both $\Zr{4}$ and $\Zn{N}$ symmetries,
the non-zero Vacuum Expectation Value (VEV) of it will break the discrete symmetries.
Hence, there can be R-parity Violation (RPV) due to the spontaneous breaking of the $\Zr{4}$ symmetry,
if the PQ field carries odd R-charge.
In this case we consider two viable scenarios of the unstable LSP.
One is that the RPV effect is so suppressed that the lifetime of the LSP
is much longer than the age of universe.
In this case, the low-energy R-parity is accidental, but is high quality such that the LSP contributes to the DM.
The other scenario is that the RPV is so large that the LSP decays
before Big Bang Nucleosynthesis (BBN), and thus the LSP is not the DM, i.e. the R-parity is low quality.
The intermediate case is excluded by experiments~\cite{Slatyer:2016qyl,
Kawasaki:1994af,Kawasaki:2004qu,Kawasaki:2008qe, Kawasaki:2017bqm}.
We will show examples for each of these scenarios.

The rest of this paper is organized as follows.
We introduce our generic model in Section~\ref{sec-model}.
Two models satisfying all the constraints are discussed in Section~\ref{sec-mex}.
We conclude this paper in Section~\ref{sec-disc}.
The Higgs potential in our non-minimal model with an extra bi-doublet field is discussed in Appendix~\ref{sec-hhbar}.
The sizes of coupling constants of operators in our examples are listed in Appendix~\ref{sec-oplist}.

\section{Generic Model}
\label{sec-model}

We introduce SUSY models based on the Pati-Salam gauge symmetry,
which is broken down to the SM at the GUT scale,
and the discrete $\ZR{4}$ and $\Zn{N}$ symmetries.
The goal of the present paper is to study the conditions under which
\renewcommand{\labelenumi}{(\arabic{enumi})}
\begin{enumerate}
\item{the mixed anomalies of the discrete/PS symmetries $\ZR{4} \times G_\PS^2$ and $\Zn{N} \times G_\PS^2$ cancel.}
\item{the anomalous $\UPQ$ symmetry is realized accidentally
         and is so high quality that it solves the strong CP problem. }
\item{the $\mu$/$b$-term are generated around the SUSY breaking scale.}
\item{any particles not in the MSSM are sufficiently heavy and quickly decaying.}
\item{the gauge coupling constants are unified at a high scale.}
\end{enumerate}
The conditions (1)  and (2) are for the non-anomalous discrete symmetry explanation for the strong CP problem.
The conditions (3) and (4) are phenomenological requirements.
The condition (5) may not be necessary for the Pati-Salam unification,
but the gauge coupling unification allows us to interpret this model as
the 4 dimensional theory resulting from an orbifold GUT
in higher dimensions~\cite{Kobayashi:2004ud,Kobayashi:2004ya}.
In this paper,
the three gauge couplings are assumed to be approximately equal up to threshold corrections at the GUT scale.

The R-parity may also arise accidentally,
and thus the LSP may decay through interactions induced by higher-dimensional operators.
Phenomenologically viable scenarios are which
\begin{itemize}
 \item[(a)] {the lifetime of the LSP is much longer than the age of universe,}
 \item[(b)] {the LSP decays before BBN.}
\end{itemize}
We will show an example for each case in Section~\ref{sec-mex}.

We will consider the following superpotential,
\begin{align}
 W = W_\PS + W_\PQ + \Delta W.
\end{align}
Here, $W_\PS$ is the leading superpotential
including the MSSM fields and fields responsible for the PS breaking.
$W_\PQ$ is the leading superpotential for the spontaneous PQ breaking sector.
$\Delta W$ includes higher-dimensional operators
which will induce the $\mu/b$-term for the Higgs doublets
as well as explicit PQ breaking and/or RPV.
$W_\PS$ and $W_\PQ$ will be introduced in Sections~\ref{sec-PSsector} and~\ref{sec-PQsector},
respectively.
The explicit PQ breaking and RPV are respectively discussed in Sections~\ref{sec-PQV} and~\ref{sec-BLV}.

\subsection{Pati-Salam sector}
\label{sec-PSsector}

\begin{table}[t]
\centering
\caption{\label{tab-matterPS}
Matter content of the generic model.
There are $N_g = 3$ generations of $Q$ and $Q^c$.
The fields $\ol{\Sigma}$ and $\ol{\Hcal}$ are not included in our minimal model.
}
\begin{tabular}[t]{c|ccc|cccc|cc}\hline
                & $\Hcal$   & $Q$ &    $Q^c$        & $X$ &   $S^c$&    $\ol{S}^c$ & $\Sigma$ & $\ol{\Sigma}$ & $\ol{\Hcal}$  \\ \hline\hline
$SU(4)_C$&  $\rep{1}$     &$\rep{4}$&  $\rep{\ol{4}}$ & $\id$ &$\rep{\ol{4}}$ & $\rep{4}$ & $\rep{6}$ & $\rep{6} $& $\rep{1}$  \\
$SU(2)_L$&  $\rep{2}$   & $\rep{2}$&  ${\id}$ & $\id$ & $\id$    & ${\id}$& $\id$&$\id$ &$\rep{2}$  \\
$SU(2)_R$&    $\rep{2}$ &   $\id$& $  \rep{\ol{2}}$     &$\id$  & $\rep{\ol{2}}$   &  $\rep{2}$    & $\id$& $\id$& $\rep{2}$  \\
\hline
$\Zbbm_{4R}$&${0}$&  ${1}$&${1}$      & $2$       &  0      &  0    & 2 & 2 & $0$      \\
$\Zbbm_{N}$ &$h$&   $-h-s$ &$s$            & 0 & $s$     &  $-s$   & $-2s$ & $2s$ & $-h$ \\
\hline
\end{tabular}
\vspace{0.5cm}
\end{table}

The matter content of the generic Pati-Salam model is shown in Table~\ref{tab-matterPS}.
$\Lambda$ is a cut-off scale for the model.
The leading superpotential is given schematically by
\begin{align}
\label{eq-superPS}
 W_\mathrm{PS} = &\   Q \Hcal Q^c + \frac{1}{2\Lambda} \ol{S}^c Q^c  \ol{S}^c Q^c    \\ \notag
                      &\quad  + X \left(\ol{S}^c S^c + \ol{\Sigma}\Sigma + \ol{\Hcal} \Hcal - v_{PS}^2\right)  + X^3
                       + S^c \Sigma S^c  + \ol{S}^c \ol{\Sigma} \ol{S}^c  + W_{s=0},
\end{align}
where
\begin{align}
\label{eq-Wszero}
W_{s = 0} = \ol{S}^c \Sigma \ol{S}^c + S^c \ol{\Sigma} S^c,
\end{align}
is allowed only if $4s\equiv 0$ modulo $N$.
The superpotential has R-charge $2$ under $\Zr{4}$, while it is neutral under $\Zn{N}$.
Throughout this paper, we omit coupling constants which may be $\order{1}$~\footnote{
The hierarchy in the SM Yukawa couplings may be explained by the Froggatt-Nielsen mechanism~\cite{Froggatt:1978nt},
as studied in Refs.~\cite{Bryant:2016tzg,Poh:2017xvg}.
}.
The MSSM quarks and leptons are contained in $Q$ and $Q^c$ as
\begin{align}
Q =
\begin{pmatrix}
 q & \ell
\end{pmatrix},
\quad
Q^c =
\begin{pmatrix}
  u^c & \nu^c \\
  d^c & e^c
\end{pmatrix},
\end{align}
where the rows are $SU(2)_R$ space and columns are the $SU(4)_C$ space.
Here, the flavor indices are implicit.
In the minimal model without $\ol{\Hcal}$, the MSSM Higgs doublets are in the bi-doublet $\Hcal$.
There are four Higgs doublets in the non-minimal model,
and two linear combinations of them correspond to the MSSM-like Higgs doublets,
see Appendix~\ref{sec-hhbar} for more details.
$S^c$, $\ol{S}^c$ are the PS breaking fields
whose VEVs are given by
$\vev{S^c} =v_{PS} \delta^{4\alpha} \delta^{i1}$ and
$\vev{\ol{S}^c} = v_{PS} \delta_{4\alpha} \delta_{i1}$,
where $\alpha$ is the $SU(4)_C$ index and $i$ is the $SU(2)_R$ index.
The Majorana masses of the right-handed neutrinos are generated
from the last term on the first line in Eq.~\eqref{eq-superPS} after PS breaking~\footnote{
This term can be obtained by integrating out a gauge singlet field $N$,
which carries charges $(1,0)$ under ($\Zr{4}$,  $\Zn{N}$),
from a renormalizable superpotential $\ol{S}^c Q N + \frac{1}{2} M_N N N$.
The mass parameter $M_N$ is expected to be $\order{\Lambda}$.
}.
The right-handed neutrino mass, $M_R$, is then given by
\begin{align}
\label{eq-MR}
 M_R \sim \frac{v_\PS^2}{\Lambda} = 10^{14}~\mathrm{GeV}\times
                            \left(\frac{v_\PS}{10^{16}~\mathrm{GeV}}\right)^2
                            \left(\frac{10^{18}~\mathrm{GeV}}{\Lambda}\right).
\end{align}
A singlet $X$ with R-charge 2 is necessary to break the PS symmetry by the F-term potential.
The vacuum which breaks PS down to the SM gauge symmetry, with
$\vev{\ol{S}^c S^c } \ne 0$ and $\vev{\ol{\Sigma}\Sigma} = \vev{\ol{\Hcal} \Hcal} = \vev{X^2} = 0$,
 is a global minimum of the scalar potential in global SUSY.
The other directions would be stabilized by
e.g. Planck suppressed operators in the K\"{a}hler potential and/or SUSY breaking mass terms.
The sextet $\Sigma$ forms a mass term with the color anti-triplet in $S^c$
(and triplet in $\ol{S}^c$ if $4s\equiv0$).
The charges under $\Zn{N}$ are chosen to be consistent with the superpotential Eq.~\eqref{eq-superPS}~\footnote{It has been shown that this superpotential is consistent with SUSY hybrid inflation~\cite{Bryant:2016tzg,Lazarides:2020zof}.
The PS breaking fields $S^c$ and $\ol{S}^c$ play a role of the waterfall fields,
so that the PS symmetry is broken during the inflation. Hence, the PS monopole is diluted away.
}.

$\ol{\Sigma}$ and $\ol{\Hcal}$ are not included in the minimal model,
but are necessary to have sizable RPV interactions consistent with the conditions (1)-(5)
as will be discussed in Section~\ref{sec-rpv}.
Without those fields, $4s\equiv 0$ modulo $N$ is required to make the triplets in both $S^c$ and $\ol{S}^c$
having masses of $\order{v_\mathrm{PS}}$ via $W_{s=0}$.
In the model with $4s\not\equiv 0$, one of the two color triplets in $\Sigma$ and $\ol{\Sigma}$
are remain massless after the PS breaking.
The light (anti-)triplet $\sigma$ ($\ol{\sigma}$) are defined as
\begin{align}
\ol{\sigma}^a :=  \eps^{abc} \Sigma_{bc},\quad \sigma_a := \eps_{abc} \ol{\Sigma}^{bc},
\end{align}
where $a,b,c=1,2,3$ are the color indices.
The (anti-)triplets $(\ol{\sigma}, \sigma)$ have hypercharge $(1/3, -1/3)$,
and thus form a vector-like pair.
In this paper, the hypercharge is defined as $Y = (B-L)/2 + T_{3R}$,
where $T_{3R}$ is a generator of $SU(2)_R$
whose eigenvalue is $0$ for a singlet, $\pm 1/2$ for a doublet.
In the non-minimal model, the bi-doublets $\Hcal$, $\ol{\Hcal}$
and the triplets will have mass via
\begin{align}
 \Delta W \supset w_0 \left( \ol{\Sigma}\Sigma +\ol{\Hcal}\Hcal \right),
\end{align}
where $w_0$ has charge $(2,0)$ and its size is expected to be the SUSY breaking scale.
In general, the VEV of the superpotential in a hidden sector would be a source for $w_0$~\cite{Casas:1992mk}~\footnote{
The mass term $w_0$ could also originate from $\vev{X}$.
}.
It is remarkable that the vector-like triplets ($\sigma$, $\ol{\sigma}$) and doublets in $\ol{\Hcal}$
can be embedded into a vector-like pair of ($\rep{5}$, $\rep{\ol{5}}$) under $SU(5)$.
Hence, gauge coupling unification may still hold even with the exotic triplets,
if they do not have any other mass larger than $\order{w_0}$
which could, in principle, originate from PQ breaking.

\subsection{Peccei-Quinn sector}
\label{sec-PQsector}

\newcolumntype{C}[1]{>{\centering\arraybackslash}p{#1}}
\begin{table}[t]
\centering
\caption{\label{tab-VLfundamental}
Charges of the PQ fields $P$, $\ol{P}$ and vector-like fields.
The field $\ol{P}$ is not included in our minimal model.
}
\begin{tabular}[t]{c|C{1.7cm}C{1.7cm}C{1.7cm}C{1.7cm}|cc}\hline
                   & $\ol{\Psi}$ & $\Psi$ & $\Psi^c$ & $\ol{\Psi}^c$ & $P$    & $\ol{P}$   \\ \hline\hline
$SU(4)_C$&  $\rep{\ol{4}} $&$\rep{4}$ & $\rep{\ol{4}} $&$\rep{4}$ & $\id$ & $\id$ \\
$SU(2)_L$&   $\rep{\ol{2}} $&$\rep{2} $  & $\rep{1} $&$\id$ & $\rep{1} $ & $\id$ \\
$SU(2)_R$&  $\id $ & $\rep{1} $&$\rep{\ol{2}}$ &$\rep{2} $&$\id$ & $\id$ \\
\hline
$\Zbbm_{4R}$& $2-  r-r_\Psi$  &$r_\Psi$ & $\ol{r}_{\Psi}$ & $2-  \ol{r}-\ol{r}_{\Psi}$ & $r$ & $\ol{r}$\\
$\Zbbm_{N}$  & $ -  p-p_\Psi$ & $p_\Psi$& $\ol{p}_{\Psi}$ &$-  \ol{p}-\ol{p}_{\Psi}$ &  $p$& $\ol{p}$  \\
\hline
\end{tabular}
\end{table}

We introduce the PQ fields, $P$ and $\ol{P}$,
which carry $\Zr{4}$ and $\Zn{N}$ charge ($r$, $p$) and ($\ol{r}$, $\ol{p}$), respectively,
and the vector-like PS quarks in Table~\ref{tab-VLfundamental}.
In the minimal model, $\ol{P}$ is not necessary,
but is required for sizable RPV interactions in addition to $\ol{\Sigma}$ and $\ol{\Hcal}$.
We will consider the KSVZ axion model with the vector-like fields~\cite{Kim:1979if,Shifman:1979if}.
The superpotential is given by
\begin{align}
\label{eq-superPQ}
W_\PQ = P \ol{\Psi} \Psi + \ol{P} ~\ol{\Psi}^c  \Psi^c + W_{\mathrm{dec}},
\end{align}
where $W_\mathrm{dec}$ contains interactions for decays of the vector-like fields.
In the minimal model without $\ol{P}$,
the second term is replaced by $P\ol{\Psi}^c \Psi^c$
and the charge of $\ol{\Psi}^c$ is given by ($2-r-\ol{r}_\Psi$, $-p-\ol{p}_\Psi$).
Since we assume that the vector-like fields have the same gauge quantum number under the PS symmetry,
$W_{\mathrm{dec}}$ will have Yukawa interactions similar to $Q\Hcal Q^c$,
depending on the charges.
In order to preserve gauge coupling unification,
we introduce $N_\Psi$ pairs of ($\Psi$, $\ol{\Psi}$) and ($\Psi^c$, $\ol{\Psi}^c$).

In this paper, we consider that the VEV of $P$ generated
by the radiatively corrected soft SUSY breaking mass term, $m_P^2 \abs{P}^2$~\cite{Moxhay:1984am}~\footnote{
We could also consider PQ breaking by the tree-level superpotential, $X (P\ol{P}-f_\PQ^2)$.
However, this may cause a fine-tuning problem, i.e. $f_\PQ \ll v_\PS$,
since we have already introduced the term $X~v_\PS^2 $ for PS breaking.}
The soft SUSY breaking mass squared will be driven to negative values
by renormalization group running due to the Yukawa coupling in Eq.~\eqref{eq-superPQ},
so that the non-zero VEV of $P$ is generated by dimensional transmutation~\cite{Coleman:1973jx}.
We expect the following form of the scalar potential,
\begin{align}
 V_P = m_P^2 \abs{P}^2 \left(\log\frac{\abs{P}^2}{f_\PQ^2} - 1\right).
\end{align}
The minimum of this potential is at $f_\PQ$
whose scale can be within the so-called axion window, $10^{9}$-$10^{12}$ GeV,
where the QCD axion can be the DM.
After PQ breaking,
$\Zr{4}$ symmetry (and simultaneously R-parity) is completely broken if $r=\pm1$, while R-parity remains unbroken if $r=2$.
The $\Zn{N}$ symmetry is broken at the PS scale, $v_\PS$, if $s\ne 0$,
while it is broken at the PQ breaking scale, $f_\PQ$, if $s=0$.
The same discussion can be applied for the other PQ field $\ol{P}$.
In this paper, we assume that the VEVs of the PQ fields are the same scale,
i.e. $\vev{P}\sim \vev{\ol{P}} \sim  f_\PQ$.

In the non-minimal model,
there should be $\mu_{\Hcal} \Hcal^2$ so that $\tan\beta \neq \infty$,
although the higgsino masses are explained by $w_0 \Hcal \ol{\Hcal}$,
see Appendix~\ref{sec-hhbar} for more details of the Higgs sector with the extra bi-doublet $\ol{\Hcal}$.
This term can be explained by the Kim-Nilles mechanism~\cite{Kim:1983dt}
if there is a term
\begin{align}
\label{eq-KNmu}
\Delta W \supset  \frac{1}{\Lambda} (P,\ol{P})^2 \Hcal^2,
\end{align}
where $(P,\ol{P})^2 = \{P^2, P\ol{P}, \ol{P}^2 \}$.
This term induces the effective $\mu$-term for $\Hcal^2$,
\begin{align}
 \mu_{\Hcal} \sim \frac{f_\PQ^2}{\Lambda}
\sim 100~\mathrm{GeV}\times \left( \frac{f_\PQ}{10^{10}~\mathrm{GeV}}\right)^2
                                                \left( \frac{10^{18}~\mathrm{GeV}}{\Lambda}\right).
\end{align}
The mass terms ($b$-terms) for the non-SM Higgs bosons are generated by the SUSY breaking.

\subsection{Anomaly cancellation}
\label{sec-anom}

We denote coefficients of the mixed anomaly of $\Zbbm^R_{4}$ to $SU(4)_C^2$, $SU(2)_L^2$ and $SU(2)_R^2$
by $\Acal^{4R}_{C}$, $\Acal^{4R}_{L}$ and $\Acal^{4R}_{R}$, respectively.
Those of $\Zbbm_{N}$ are denoted by $\Acal^{N}_{C}$, $\Acal^{N}_{L}$ and $\Acal^{N}_{R}$.
The coefficients are given by~\cite{Ibanez:1991pr,Ibanez:1991hv,Ibanez:1992ji,Dreiner:2005rd,Araki:2007zza,Araki:2008ek},
\begin{align}
\label{eq-anom4R}
 \Acal_C^{4R} \equiv  1 + N_{\ol{\Sigma}} -(r+\ol{r}) N_\Psi,\quad
 \Acal_L^{4R} \equiv   1 - N_{\ol{\Hcal}} -2rN_\Psi,\quad
 \Acal_R^{4R} \equiv  1 - N_{\ol{\Hcal}} -2\ol{r} N_\Psi,
\end{align}
modulo $2$ and
\begin{align}
 \Acal_C^{N} =&\  - 2s(1 - N_{\ol{\Sigma}}) - h N_g- (p+\ol{p})N_\Psi,  \notag \\
 \Acal_L^{N} =&\   h(1-N_{\ol{\Hcal}}) - 2(h+s) N_g- 2 p N_\Psi, \\
 \Acal_R^{N} =&\   h(1-N_{\ol{\Hcal}}) + 2s N_g - 2 \ol{p} N_\Psi, \notag
\end{align}
where $N_g = 3$ is the number of generations of SM fermions.
Here, $N_{\ol{\Hcal}}$, $N_{\ol{\Sigma}}$
are respectively the number of $\ol{\Hcal}$, $\ol{\Sigma}$, while we take $N_{\Hcal} = N_{\Sigma} = 1$.
In the minimal model, $N_{\ol{\Hcal}} = N_{\ol{\Sigma}} = 0$,
$\ol{r} = r$ and $\ol{p}=p$.

The conditions for anomaly cancellation are given by
\begin{align}
 \Acal_C^{4R} \equiv \Acal_L^{4R} \equiv \Acal_R^{4R} \quad&\ \mathrm{modulo}~2,  \\
 \Acal_C^{N} \equiv \Acal_L^{N} \equiv \Acal_R^{N} \quad&\ \mathrm{modulo}~N.
\end{align}
The anomaly is completely canceled if these are vanishing,
while these can be canceled by the Green-Schwartz mechanism~\cite{Green:1984sg}
if these are non-vanishing but have a universal value.
In this paper, we will consider the minimal case $N_{\ol{\Hcal}} = N_{\ol{\Sigma}} = 0$
and the next-to-minimal case $N_{\ol{\Hcal}} = N_{\ol{\Sigma}} = 1$.

\subsection{The axion quality}
\label{sec-PQV}

There will be numerous higher dimensional operators which may explicitly break the PQ symmetry.
In general, the $\theta$ angle will be shifted at tree-level by a PQ breaking term in the superpotential,
\begin{align}
 W_{\cancel{\PQ}} \supset \frac{1}{\Lambda^{k+2l+m+n-3}}w_0^k \Hcal^{2l} P^{m} \ol{P}^{n},
\end{align}
where $k,l,m,n$ are integers.
Note, contributions from operators with $\ol{\Hcal}^2$ cannot be the leading ones, since as long
as $\Delta W \supset (P,\ol{P})^2 \Hcal^2/\Lambda$ in Eq.~\eqref{eq-KNmu}
is allowed for the $\mu/b$-term and we have $ \vev{\ol{\Hcal}}^2 < w_0 f_\PQ^2/\Lambda$,
which is satisfied for the typical values of VEVs.
The powers satisfy
\begin{align}
  2k+mr+n\ol{r} \equiv 2\quad&\mathrm{modulo}~4, \\
  2lh+mp+n\ol{p}   \equiv 0\quad&\mathrm{modulo}~N.
\end{align}
This term will affect the $\theta$ angle via the F-term potential and the soft SUSY breaking $A$-term.
The leading PQ breaking in the F-term potential will be an interference term
between the PQ conserving and breaking terms in the superpotential.  
Consider that the F-term VEVs will be
\begin{align}
 F_P,~F_{\ol{P}} \lesssim w_0^2,\quad  F_\Hcal \lesssim v_H w_0,
\end{align}
where $F_\Phi$ is an F-term of a superfield $\Phi = P,\ol{P}, \Hcal$. 
The F-term of the Higgs field will depend on $v_H$ because $\Hcal^2$ is a gauge singlet combination.
We can show that
\begin{align}
 V_F \supset \sum_{\Phi=P,\ol{P},\Hcal} F_\Phi \frac{\partial W_{\cancel{\PQ}}}{\partial \Phi}
              \lesssim  w_0 W_{\cancel{\PQ}},
\end{align}
where the right-hand side corresponds to the A-term contribution. 
Hence, it is enough to confirm that $\Delta\theta$ from the A-term,
\begin{align}
\label{eq-PQVW0}
 \Delta \theta \sim&\ \frac{w_0^{1+k} v_H^{2l} f_\PQ^{m+n}}{\LQCD^4 \Lambda^{k+2l+m+n-3}}   \\ \notag
                     \sim&\ 10^{63-13k-32l-8(m+n)} \times
                                                        \left(\frac{w_{0}}{10^5~\mathrm{GeV} }\right)^{1+k}
                                                        \left(\frac{f_\PQ}{10^{10}~\mathrm{GeV}}  \right)^{m+n}
                                                        \left(\frac{10^{18}~\mathrm{GeV}}{\Lambda}  \right)^{k+2l+m+n-3},
\end{align}
 is sufficiently suppressed.

We set the Higgs VEV, $v_H = 100$ GeV
and the QCD scale in front of the axion potential, $\LQCD = 100$ MeV.
For $k=l=0$, $m+n \ge 10$ is typically required for $\Delta \theta  < 10^{-10}$
to solve the strong CP problem.
Clearly, $\Zr{4}$ alone cannot suppress the self-coupling of $P$ up to this order,
and an additional symmetry such as the $\Zn{N}$ symmetry is necessary.
Note that the PS breaking VEV cannot be the leading PQ breaking effect,
since $\ol{S}^c S^c$ is a unique gauge singlet combination whose VEV is non-zero,
but this is neutral under the discrete symmetries.
Hence, $\vev{P}\sim\vev{\ol{P}} \sim f_\PQ$ will be the largest source of PQ breaking.

There might also be radiative corrections to the potential induced by explicit PQ breaking interactions.
Let us consider a term in a superpotential,
\begin{align}
 W_{\cancel{\PQ}} \supset \frac{w_0^a \Hcal^{2b} P^c \ol{P}^{d} (S^c, \ol{S}^c)^e }{\Lambda^{a+2b+c+d+e-1}} \phi_1 \phi_2,
\end{align}
where $\phi_1$ and $\phi_2$ can be any fields allowed by the discrete symmetries
and they do not need to have a non-zero VEV. Here, $a, b,c,d, e$ are integers.
The integers $a,e = 0,1$ for the leading contributions.
The PS breaking fields $S^c$ or $\ol{S}^c$
can appear in the leading contribution from e.g. $\phi_1 = \Hcal$, $\phi_2 = Q$.
There may be 1-loop corrections mediated by $\phi_1$, $\phi_2$ to the $\theta$ angle given by
\begin{align}
\label{eq-PQVW1}
\Delta \theta &\ \sim \frac{1}{16\pi^2}
             \frac{w_0^{2+2a} v_H^{4b} f^{2c+2d}_\PQ v_\PS^{2e} }{\LQCD^4  \Lambda^{2a+4b+2c+2d+2e-2} }  \\ \notag
                    \sim&\ 10^{48-26a-64b-16(c+d)-4e}  \\ \notag
                       &\             \times \left(\frac{w_{0}}{10^5~\mathrm{GeV} }\right)^{2+2a}
                                                        \left(\frac{f_\PQ}{10^{10}~\mathrm{GeV}}  \right)^{2c+2d}
                                                        \left(\frac{10^{16}~\mathrm{GeV}}{v_\PS}  \right)^{2e}
                                                        \left(\frac{10^{18}~\mathrm{GeV}}{\Lambda}  \right)^{2a+4b+2c+2d+2e-2}.
\end{align}
For $a = b =0$ and $e=1$,
$c + d \ge 4$ may be required to keep the axion quality, $\Delta \theta < 10^{-10}$.
Thus the radiative correction may not spoil the axion solution to the strong CP problem
if explicit PQ breaking terms are absent up to dimension-7 operators.

The PQ breaking in K\"{a}hler potential can also affect the $\theta$ angle.
The leading contribution to the scalar potential will be given by
\begin{align}
\label{eq-KahlerPQV}
&\ V_{\cancel{\PQ}} \supset w_0^2  K_{\cancel{\PQ}} \\ \notag
  &\   \supset    \frac{1}{\Lambda^{k+2l+m+n-2}} w_0^{k+2} \Hcal^{2l} P^m \left(\ol{P}^\dag\right)^n
       +  \frac{1}{\Lambda^{a+2b+c+d+e}} w_0^{a+2} \Hcal^{2b} P^c \left(\ol{P}^\dag\right)^d
     (S^{c(\dag)},\ol{S}^{c(\dag)})^e \phi^{(\dag)}_1 \phi_2^{(\dag)}.
\end{align}
In the minimal model, we will assume that $\Delta W \supset w_0 \Hcal^2$ is allowed for the $\mu/b$-term,
so $\Hcal^\dag$ has the same charge as $\Hcal$ and will not induce a new PQ breaking operator.
In the non-minimal model, operators involving $(\Hcal^\dag, \ol{\Hcal})^2$
cannot be the leading one for the same reason as those involving $\ol{\Hcal}^2$ in the superpotential.
Thus, the terms in Eq.~\eqref{eq-KahlerPQV} give the leading PQ violating effect from the K\"{a}hler potential.
The shift in $\theta$ from the first term is estimated as
\begin{align}
\label{eq-PQVK0}
 \Delta \theta \sim&\ \frac{w_0^{k+2} v_H^{2l} f_\PQ^{m+n}}{\LQCD^4 \Lambda^{k+2l+m+n-2}}   \\ \notag
          \sim&\ 10^{50-13k-32l-8(m+n)} \times
                                                       \left(\frac{w_{0}}{10^5~\mathrm{GeV} }\right)^{2+k}
                                                        \left(\frac{f_\PQ}{10^{10}~\mathrm{GeV}}  \right)^{m+n}
                                                        \left(\frac{10^{18}~\mathrm{GeV}}{\Lambda}  \right)^{k+2l+m+n-2}.
\end{align}
For $k = l =0$, $m+n \ge 8$ will be required for the axion quality.
The loop correction mediated by $\phi_1^{(\dag)}$ and $\phi_2^{(\dag)}$ is estimated as
\begin{align}
\label{eq-PQVK1}
 \Delta \theta \sim&\ \frac{w_0^{2a+4} v_H^{4b} f_\PQ^{2c+2d} v_\PS^{2e}}{16\pi^2 \LQCD^4 \Lambda^{2a+4b+2c+2d+2e}}  \\ \notag
       \sim&\ 10^{22-26a-64b-16(c+d)-4e}  \\ \notag &\ \times
                                                       \left(\frac{w_{0}}{10^5~\mathrm{GeV} }\right)^{2a+4}
                                                        \left(\frac{f_\PQ}{10^{10}~\mathrm{GeV}}  \right)^{2c+2d}
                                                        \left(\frac{v_\PS}{10^{16}~\mathrm{GeV}}  \right)^{2e}
                                                        \left(\frac{10^{18}~\mathrm{GeV}}{\Lambda}  \right)^{2a+4b+2c+2d}.
\end{align}
For $a=b=0$, $c+d=2$ and $e=0$ ($1$),
the shift of $\theta$ is comparable to the experimental bound, $\Delta \theta \sim 10^{-10}$ ($10^{-14}$). It is much smaller for $c+d>2$.

\subsection{R-parity violation and proton stability}
\label{sec-BLV}

Stability of the proton and LSP are not ensured in this model,
because the discrete symmetries $\Zr{4}\times\Zn{N}$ are broken by the PQ fields.
If all the PQ fields have even R-charge, R-parity remains unbroken.
The LSP will be stable,
but dimension-5 (or higher) baryon and lepton number violating operators would destabilize the proton.
Although these may be highly suppressed as discussed in Refs.~\cite{Lee:2010gv,Lee:2011dya}.
If $P$ and/or $\ol{P}$ have odd R-charge, $\Zr{4}$ symmetry is completely broken.
Then R-parity might appear as an accidental symmetry, if the RPV operators are highly suppressed.

The $\Zn{N}$ symmetry is also completely broken by $f_\PQ$ if
the charge of $P$ and/or $\ol{P}$ is not a divisor of $N$.
In this model, the dimension-4 RPV operators are induced from the operators,
\begin{align}
\label{eq-RPV}
 Q\Hcal S^c: (1,0),\quad Q\ol{\Hcal} S^c: (1,-2h),\quad  QQ^cQS^c:(3,-2h),\quad Q^cQ^cQ^cS^c: (3,4s),
\end{align}
after the PS breaking.
Here, the values in the parentheses are charges under ($\Zr{4}$, $\Zn{N}$).
The first three operators violate lepton number, while the last one violates baryon number.
The superpotential is given by
\begin{align}
\label{eq-Wrpv}
 W_{\mathrm{RPV}} =  y_{L} Q\Hcal S^c +
                                     \ol{y}_L Q\ol{\Hcal} S^c +
                                     \ka_L QQ^c Q S^c +
                                      \ka_B Q^c Q^c Q^c S^c,
\end{align}
where the coupling constants depend on
certain combinations of $P$, $\ol{P}$, $w_0$ and $\Hcal^2$ to be consistent with the discrete symmetry.
The effective $\Delta L=1$ ($\Delta B=1$) Yukawa coupling $\la_L$ ($\la_B$)
induced by those operators are given by
\begin{align}
\la_{L} \sim \max\left( \frac{y_L v_\PS}{w_0},  \frac{\ol{y}_L v_\PS}{w_0}, \ka_L v_\PS \right),
\quad
\la_{B} \sim  \ka_B v_\PS.
\end{align}
Here, we assume that the Yukawa coupling of $Q\Hcal Q^c$ is $\order{1}$
and the higgsino mass is $\order{w_0}$.
The bilinear $\Delta L =1$ Yukawa couplings are induced by rotating away the bi-linear RPV terms
by redefining Higgs and leptons.
The proton decay constraints on the RPV operators are
\begin{align}
\la_L \la_B \lesssim 10^{-27} \times \left(\frac{m_{\tilde{f}}}{1~\mathrm{TeV}}\right)^2,
\end{align}
where $m_{\tilde{f}}$ is a sfermion mass.

The LSP will become unstable if there are sizable RPV effects.
If the lepton number violation dominates RPV,
the lifetime of a neutralino LSP via the three-body decay, $\chi \to \nu \ell \ell^\dag$,
where $\ell \ell^\dag = e^+ e^-, \nu \ol{\nu}$,
is estimated as~\cite{Berezinsky:1996pb,Barbier:2004ez},
\begin{align}
\label{eq-tauchiL}
 \tau_{\chi} \sim&\ \frac{1536\pi^3}{ g_{\nu Z}^4 N_{\chi\nu}^2 }
                                                    \frac{m_{\mathrm{Z}}^4}{m^5_{\chi}}
                       \sim  1~\mathrm{s}\times \left(\frac{0.1}{g_{\nu Z}} \right)^4
                                         \left(\frac{10^{-14}}{ N_{\chi\nu} }\right)^2
                                        \left(\frac{10^{4}~\mathrm{GeV}}{m_{\chi}} \right)^5.
\end{align}
Here, we consider the three-body decay through a Z-boson whose mass is set at 100 GeV~\footnote{
The decay through a W-boson will have the same size.}.
The decays through the EW boson dominate over those through sfermions,
if the sfermions are heavier than $\order{10~\mathrm{TeV}}$, see Eq.~\eqref{eq-tauchiB}.
$g_{\nu Z}$ is the coupling constant of the neutrinos to a Z-boson.
The mass of the neutralino LSP is denoted by $m_\chi$.
$N_{\chi\nu}$ is the mixing angle of the LSP and the neutrino whose size is estimated as
\begin{align}
\label{eq-nhbl}
 N_{\chi \nu}
    \sim  N_{\chi \tilde{H}} \times
    \max\left( \frac{y_L v_\PS}{w_0},  \frac{\ol{y}_L v_\PS}{w_0}, \frac{\ka_L v_\PS}{16\pi^2} \right),
\end{align}
where $N_{\chi \tilde{H}}$ is the fraction of higgsinos in the lightest neutralino $\chi$.
The last one comes from mixing at the 1-loop level via the RPV Yukawa coupling.
Here, the MSSM Yukawa coupling is set to $1$ for simplicity.
The LSP decays before BBN, i.e. $\tau_\chi \lesssim 1~\mathrm{s}$,
if $N_{\chi\nu} \gtrsim 10^{-14}$.
On the other hand, the lifetime is longer than $\order{10^{24}~\mathrm{s}}$
if $N_{\chi\nu} \lesssim 10^{-26}$, such that the neutralino LSP is the stable DM and its decay
does not affect the cosmic microwave background~\cite{Slatyer:2016qyl}.

If the baryon number violation is the dominant one, then the higgsino-lepton mixing is negligible.
Hence, the neutralino decays through squarks and its lifetime is estimated as
\begin{align}
\label{eq-tauchiB}
 \tau_{\chi} \sim&\ \frac{1536\pi^3}{ g_{\chi\tilde{q}}^2 \la_{B}^2 }
                                                    \frac{m_{\tilde{q}}^4}{m^5_{\chi}}
                       \sim  1~\mathrm{s}\times \left(\frac{0.1}{g_{\chi\tilde{q}}} \right)^2
                                         \left(\frac{10^{-9}}{\la_{B}} \right)^2
                                        \left(\frac{m_{\tilde{q}}}{10^{5}~\mathrm{GeV}}\right)^4
                                        \left(\frac{10^{4}~\mathrm{GeV}}{m_{\chi}} \right)^5,
\end{align}
where $g_{\chi \tilde{q}}$ is a coupling constant for the quark-squark-LSP interaction
and $m_{\tilde{q}}$ is a squark mass.
The LSP decays before BBN if $\la_{\mathrm{B}} \gtrsim 10^{-9}$,
while the LSP is a stable and invisible DM particle if $\la_{\mathrm{B}} \lesssim 10^{-21}$.

Proton decay may also be mediated by dimension-5 operators,
\begin{align}
\label{eq-DimFive}
 QQQQ: (0,-4h-4s),\quad Q^cQ^cQ^cQ^c: (0, 4s),
\end{align}
in the superpotential and/or dimension-6 operator,
\begin{align}
 Q^\dag Q^\dag Q^c Q^c: (0, 2h+4s),
\end{align}
 in the K\"{a}hler potential.
The effective cut-off scale for the dimension-5 and -6 operators, $\Lambda_5$ and $\Lambda_6$,
should be larger than $\order{10^{27}}$ and $\order{10^{15}}$ GeV, respectively.
These operators will be sufficiently suppressed by the discrete symmetries. Note, however, that the dimension-5 and -6 operators, as well as the RPV operators, might also be generated after integrating out vector-like triplets which are much lighter than the GUT scale.  Clearly we need
to check that these are also suppressed.

Let us consider the vector-like triplets $(T, \ol{T})$,
which are in $(\Psi, \ol{\Psi})$, $(\ol{\Psi}^c, \Psi^c)$ or $(\ol{\Sigma}, \Sigma)$,
with a superpotential,
\begin{align}
\label{eq-VLtriplet}
 W \supset &\ m_T \ol{T} T
        + T       \left( \mu_{\ol{T}} \ol{\Qcal}^T_1 + \la_{\ol{T}} \ol{\Qcal}^T_2 + {\ka}_{\ol{T}} \ol{\Qcal}^T_3   \right)
       + \ol{T} \left( \mu_T \Qcal^T_1 + \la_T \Qcal^T_2 + \ka_T\Qcal^T_3   \right),
\end{align}
where $\Qcal_k^T$ and $\ol{\Qcal}_k^T$ are composed
of $\{Q, Q^c, \Sig, \oSig \}$ for $(T,\ol{T}) =(\Psi_q, \ol{\Psi}_q)$, $(\ol{\Psi}^c_q, \Psi^c_q)$,
and $\{Q, Q^c\}$ for $(T,\ol{T}) =( \sigma, \ol{\sigma})$.
Here, integer $k$ represents the mass dimension of $\Qcal_k^T$ and $\ol{\Qcal}^T_k$.
$\Psi_q$ ($\ol{\Psi}_q$) and $\ol{\Psi}^c_q$ ($\Psi^c_q$) are color (anti-)triplets 
in $\Psi$ ($\ol{\Psi}$) and $\ol{\Psi}^c$ ($\Psi^c$), respectively.  
The coupling constants depend on the non-zero VEVs of the fields.
After integrating out the vector-like triplets,  we have
\begin{align}
 W\supset  \frac{1}{m_T}  &
           \left( \mu_{T} \mu_{\ol{T}}  Q_1^T \ol{Q}_1^{T}
                  +     \mu_{T} \la_{\ol{T}} \Qcal_1^T \ol{\Qcal}_2^T
                    +  \la_{T} \mu_{\ol{T}} \Qcal_2^T \ol{\Qcal}_1^T \right. \\ \notag
                &\ \left. \quad  + \mu_T \kappa_{\ol{T}} \Qcal_1^T \ol{\Qcal}_3^T
                    +  \la_T \la_{\ol{T}} \Qcal_2^T \ol{\Qcal}_2^T
                   + \ka_T \la_{\ol{T}} \Qcal_3^T \ol{\Qcal}_1^T  \right), 
\end{align}  where we omit the higher-dimensional operators.
The mass mixing effects of the first term may be sufficiently suppressed
as will be shown in explicit examples in Section~\ref{sec-mex}.
The latter two terms in the first line may induce the RPV Yukawa couplings
and the second line may include the dimension-5 operators.
Without discussing details, the proton will be stable if
\begin{align}
 \max\left(\frac{\mu_T \ola_{T}}{m_T},  \frac{ \la_T \ol{\mu}_{T}}{m_T} \right)
&\ \ll 10^{-17} \times \left(\frac{10^{-10}}{\la_L}\right),  \\
\max \left(\frac{\mu_T \oka_{T}}{m_T},
\frac{\la_T \ola_{T}}{m_T},   \frac{\ka_T \ol{\mu}_{T}}{m_T} \right)
&\ \ll \left(10^{27}~\mathrm{GeV}\right)^{-1}.
\end{align}
Since we will find $\mu_T/m_T \lesssim 1$ in our examples,
$\max(\la_T, \la_{\ol{T}}) \ll 10^{-17}$ and $\max(\ka_T, \ka_{\ol{T}}) \ll 10^{-27}~\mathrm{GeV}^{-1}$
are sufficient conditions for the proton stability.

In the K\"{a}hler potential,
the sizable dimension-6 operator could be induced by integrating out the color triplets from
\begin{align}
 K \supset&\
 \frac{1}{\Lambda}  (Q^c Q^c)^\dag \left(\zeta_\Sigma \Sigma +  \frac{\zeta_\Psi  }{\Lambda} S^c \Psi^c\right)
            + \frac{1}{\Lambda}  \left( \zeta^c_{{\Sigma}} \ol{\Sigma}
                    +  \frac{\zeta^c_\Psi }{\Lambda} \ol{S}^c \ol{\Psi}^c  \right)^\dag   QQ,   
\end{align}
where $\zeta_\Phi^{(c)}$, $\Phi = \Sigma, \Psi, \Psi^c$
are coupling constants implicitly depending on the non-zero VEVs of gauge singlet combinations.
The dimension-6 operators will be generated
by integrating out scalar components in the vector-like fields,
together with Yukawa couplings,
\begin{align}
 W &\ \supset \la_{Q^c {\Sigma} Q^c} Q^c \Sigma Q^c   + \la_{Q \ol{\Sigma} Q} Q \ol{\Sigma} Q
                        + \ka_{Q^cQ^c {S}^c \Psi^c} Q^cQ^c {S}^c \Psi^c
                                    +  \ka_{QQ \ol{S}^c \ol{\Psi}^c} QQ \ol{S}^c \ol{\Psi}^c.
\end{align}
Note that $\la_{Q^c \Sigma Q^c}$ ($\la_{Q \ol{\Sigma} Q}$)
is a part of $\la_{\ol{\sigma}}$ ($\la_{\sigma}$) in Eq.~\eqref{eq-VLtriplet}
and
$v_\PS \ka_{Q^cQ^c S^c \Psi^c}$ ($v_\PS \ka_{QQ\ol{S}^c \ol{\Psi}^c}$)
is a part of $\la_{\Psi^c_q}$ ($\la_{\ol{\Psi}^c_q}$).
The dimension-6 operators arise as
\begin{align}
         \int d^4\theta K
           \supset &\
  \frac{1}{\Lambda}
     (Q^c Q^c)^\dag \left(\zeta_\Sigma F_{\ol{\sigma}} +  \frac{\zeta_\Psi  }{\Lambda} v_\PS F_{\Psi^c_q}\right)
            + \frac{1}{\Lambda}  \left( \zeta^c_{\Sigma} F_{{\sigma}}
                    +  \frac{\zeta^c_\Psi }{\Lambda} v_{\PS} F_{\ol{\Psi}^c_q}
                   \right)^\dag   QQ  \\
\label{eq-dim6form}
 \to&\
  \left(\frac{\zeta_\Sigma \la_{Q\ol{\Sigma} Q}}{m_\sigma\Lambda}
       + \frac{\zeta_\Psi v_\PS^2 \ka_{QQ\ol{S}^c\ol{\Psi}^c} }{m_{\Psi^c} \Lambda^2}\right)
           \left(u^c e^c \right)^\dag qq   \\ \notag
&\ \hspace{2cm} +   \left(\frac{\zeta^c_\Sigma \la_{Q^c {\Sigma} Q^c}}{m_\sigma\Lambda}
       + \frac{\zeta^c_\Psi v_\PS^2 \ka_{Q^cQ^cS^c\Psi^c} } {m_{\Psi^c} \Lambda^2}\right)
           \left(u^c d^c \right)^\dag q\ell,  
\end{align}
where $Q, Q^c$ are fermionic components of the superfields of the same symbols.
Here, $F_\Phi$, $\Phi = \sigma, \ol{\sigma}, \Psi^c_q, \ol{\Psi}^c_q$,
are the F-terms of color (anti-)triplets in the vector-like fields.
The scalar fields in the vector-like fields are integrated out in the second equality.
For example, the F-term of ${\sigma}$ in $\ol{\Sigma}$ is given by
\begin{align}
\label{eq-Fsigma}
 F_{{\sigma}} \sim - m_{\sigma} \ol{\sigma}^*  \sim  \frac{\la_{Q^c{\Sigma} Q^c} u^c d^c}{m_\sigma}.
\end{align}
Thus, the sufficient condition is
$\max(\la_{Q^c\Sigma Q^c},  \la_{Q{\oSig}Q})  \ll 10^{-7}$ for $m_\sigma = 10^{5}$ GeV
and
$\max(v_\PS \ka_{Q^cQ^cS^c\Psi^c},  v_\PS \ka_{QQ \ol{S}^c\ol{\Psi}^c} ) \ll 1$
for $m_\Psi = 10^{10}$~GeV, since $\zeta_{\Sigma, \Psi}^{(c)} < \order{1}$.
These are much weaker constraints than those from the superpotential.
Operators more suppressed by $\Lambda$
will always be sufficiently small because the effective cut-off scale will be larger than $\Lambda$
as far as $\mu_T < 1~\mathrm{GeV}$, $\la_T < 1$ and $\ka_T < 1~{\mathrm{GeV}^{-1}}$,
which are clearly satisfied in our examples.

These sufficient conditions in the superpotential and K\"{a}hler potential
are satisfied in our examples shown in the next section,
so we will not discuss any more details of proton decay in this paper.

\section{Model examples}
\label{sec-mex}
\subsection{Minimal model: high-quality LSP}
\label{sec-minimal}

Let us first consider the minimal model with $N_{\ol{\Hcal}} = N_{\ol{\Sigma}} = N_{\ol{P}} = 0$,
$N_\Psi = 1$ and $s=0$.
The anomaly cancellation for $\Zn{N}$ implies,
\begin{align}
 - 3 h \equiv - 5h \equiv h \quad \mathrm{modulo}~N.
\end{align}
The solution is $h=0$ if $N$ is odd, and is $h\equiv 0$ modulo $N/2$ if $N$ is even.
Under this condition, $\Hcal^2$ is neutral under the discrete symmetries.
The mass term itself is forbidden by the $\Zr{4}$ symmetry, but the $\mu$-term will be generated
after SUSY breaking by $W \supset w_0 \Hcal^2$.
In addition, the $b$-term will be generated from this term after SUSY breaking.
Therefore, the condition (3) is satisfied whenever the condition (1) is satisfied.

\begin{table}[t]
\centering
\caption{\label{tab-minimal}
The charges under the discrete symmetries and the accidental $U(1)_\PQ$ in the minimal model.
}
\begin{tabular}[t]{c|ccc|cccc|cccc|c}\hline
                & $\Hcal$   & $Q$ & $Q^c$  & $X$ &$S^c$& $\ol{S}^c$ & $\Sigma$ & $\ol{\Psi}$ & $\Psi$& $\Psi^c$&$\ol{\Psi}^c$ & $P$ \\ \hline\hline
$\Zbbm_{4R}$&${0}$&  ${1}$&${1}$   &  $2$       &  $0$      &  $0$    & $2$ & $0$ & $1$  & $1$ & $0$ & $1$    \\
$\Zbbm_{5}$ &$0$&   $0$ &$0$         & $0$    &  $0$     &  $0$   & $0$ & $4$ & $0$ & $0$  & $4$ & $1$\\
\hline
$\UPQ$ & $0$ & $0$ & $0$ & $0$   & $0$ & $0$ & $0$ & $-1$ & $0$ & $0$ & $-1$  & $1$\\
\hline
\end{tabular}
\end{table}

For concreteness, we shall choose the charges
\begin{align}
 N=5, \quad r= p = r_\Psi
= 1, \quad
h = p_{\Psi}
= 0.
\end{align}
The charges of the fields are listed in Table~\ref{tab-minimal}.
With this charge assignment, $\Psi$ ($\Psi^c$) have the same charge as $Q$ ($Q^c$),
so that these are like a fourth generation of the MSSM (s)fermions,
but with vector-like masses of $\order{f_\PQ}$, see Eq.~\eqref{eq-superPQ}.
It is clear that the Yukawa coupling $Q\Hcal Q^c$ will induce the decays of the vector-like particles,
and thus this model satisfies the condition (4).
Gauge coupling unification is preserved and condition (5) is satisfied,
since all the triplets in $S^c$, $\ol{S}^c$ and $\Sigma$ have masses of $\order{v_\PS}$.
There is a fourth family of vector-like fields with mass of $\order{f_\PQ}$
and the MSSM particles have mass less than the SUSY breaking scale.

We can find an accidental anomalous $\UPQ$ symmetry whose charges are shown
in the last row of Table~\ref{tab-minimal}.
The MSSM particles cannot carry $\UPQ$ charge to be consistent with the PS superpotential,
\begin{equation}
W_{\PS} \supset \frac{1}{\Lambda}(\ol{S}^c Q^c)^2 + Q\Hcal Q^c + w_0 \Hcal^2.
\end{equation}
Hence, only the vector-like quarks carry the $\UPQ$ charge, so the model has the KSVZ axion.
The PQ breaking superpotential is given by
\begin{align}
\label{eq-WPQVMIN}
W_{\cancel{\mathrm{PQ}}}
= \frac{P^{10}}{\Lambda^7} + \frac{P^5}{\Lambda^4} Q^c \ol{S}^c
     + \frac{P^5}{\Lambda^5} \Hcal Q S^c + \cdots.
\end{align}
These terms induce the shift in the $\theta$ angle by $\sim 10^{-17}$, $10^{-32}$ and $10^{-36}$, respectively,
and thus the PQ symmetry is so high quality that it solves the strong CP problem.
All the operators which can contribute to $\Delta\theta$ are listed in Table~\ref{tab-PQVexMIN}
of Appendix~\ref{sec-oplist}.
In this model, the axion domain-wall number is $N_{\mathrm{DW}} = 4$ and is not unity.
However, the domain-wall is unstable due to the explicit PQ breaking effects,
thus it would not cause a cosmological problem~\cite{Sikivie:1982qv}~\footnote{
It would also be solved by the dynamics of multiple scalar fields~\cite{Ibe:2019yew}.
}.

The standard R-parity violations are extremely suppressed due to the discrete symmetry.
In the minimal model, the lowest order for which $P$ can couple to the RPV operators is $P^5$,
because none of the operators in Eqs.~\eqref{eq-RPV} have $\Zn{5}$ charge.
In fact, the bilinear RPV term is of order $\sim 10^{-24}$ GeV
and the lepton number violating Yukawa couplings are of order $\sim 10^{-55}$ in this model,
and thus $\la_L  \sim 10^{-29}$ for $w_0 \sim 10^{5}$ GeV.
The coefficient of dimension-4 baryon number violating operator  is of order $10^{-55}$.
This is clearly sufficiently small to make the lifetime of the neutralino LSP
$longer$ than the age of universe,
see Eq.~\eqref{eq-nhbl}.
For the same reason, the proton lifetime is extremely long.
Therefore, R-parity exists very precisely in this model.
The full list of operators and their typical values relevant to the proton decays are shown
in Table~\ref{tab-RPVexMIN} of Appendix~\ref{sec-oplist}.
This conclusion will not be changed
since the MSSM fields do not carry $\Zn{N}$ charges as is required
by the anomaly cancellation in the minimal model.

The stable LSP may or may not be a problem.
The LSP is known to be an attractive candidate for the DM
if the neutralino masses are in the suitable range, e.g. higgsino $\sim 1$ TeV
and the conventional thermal freeze-out scenario is working~\cite{Cirelli:2005uq,Cirelli:2007xd}.
However, the LSP tends to overclose the universe in high-scale SUSY scenarios.
In particular, the non-thermal production from the gravitino and/or moduli often overproduce the LSP~\cite{Kawasaki:1994af,Endo:2006zj,Kawasaki:2008qe}.
This overproduction problem could be solved if the LSP is much lighter than the TeV scale.
An axino with a mass $\lesssim \order{\mathrm{keV}}$,
is a candidate for such a particle~\cite{Choi:2011yf,Kim:2012bb,Choi:2013lwa} if it is sufficiently stable.

We shall discuss a case of $\order{\mathrm{keV}}$ axino LSP.
The axino mixes with neutrinos by the RPV effects.
Defining the axion superfield $A$ via
\begin{align}
 P  = f_\PQ e^{A/f_\PQ},
\quad
A = \frac{1}{\sqrt{2}}  \left( s+ia \right) + \sqrt{2} \theta \tilde{a} + \theta^2 F_A,
\end{align}
where $s$, $a$ and $\tilde{a}$ are saxion, axion and axino, respectively.
$F_A$ is the F-term of the superfield $A$.
Integrating out the right-handed neutrinos, we find
\begin{align}
\label{eq-wanu}
 W \supset&\ \frac{1}{M_R} \left(\ell H_u + \frac{v_\PS P^5}{\Lambda^4}  \right)^2
                \supset
         \frac{f^4_\PQ}{v_\PS \Lambda^3}  \left( f_\PQ  H_u^0 \nu + v_H A \nu + \frac{v_H}{f_\PQ} AA\nu \right),
\end{align}
where $\order{1}$ coefficients are omitted and $M_R$ is replaced by using Eq.~\eqref{eq-MR}.
The first two terms induce the RPV higgsino-neutrino and axino-neutrino mixing, respectively.
These could affect the neutrino mass by
\begin{align}
 \delta m_{\nu} \sim&\  \left(\frac{v_H f_\PQ^4}{v_\PS \Lambda^3}  \right)^2
                                   \max\left(\frac{f_\PQ^2}{w_0},\ \frac{v_H^2}{m_{\tilde{a}}}  \right)   \\ \notag
               \sim&\        10^{-32}~\mathrm{eV} \times
                                       \left( \frac{10^{16}~\mathrm{GeV}}{v_\PS} \right)^2
                                       \left( \frac{f_\PQ}{10^{10}~\mathrm{GeV}} \right)^8
                                       \left( \frac{10^{18}~\mathrm{GeV}}{\Lambda} \right)^6
                                       \left(
                                       \frac{
                                      \max\left({f_\PQ^2}/{w_0},\ {v_H^2}/{m_{\tilde{a}}}  \right)
                                      }{10^{15}~\mathrm{GeV}}\right),
\end{align}
where $m_{\tilde{a}}$ is the axino mass and higgsino mass is assumed to be $\order{w_0}$.
Thus the mixing will not affect neutrino masses.

The axino will dominantly decay by $\tilde{a} \to \nu a$ or $\tilde{a}\to \nu \ell\ell^\dag $,
where $\ell \ell^\dag = e^+e^-$, $\nu\nu$,
as discussed in Refs.~\cite{Kim:2001sh,Hooper:2004qf,Chun:2006ss,Endo:2013si}.
The decay to electrons are allowed when $m_{\tilde{a}} > 2m_e$.
The first decay mode is induced by the last term in Eq.~\eqref{eq-wanu}
and the lifetime via this mode is estimated as,
\begin{align}
 \tau_{\tilde{a}\to \nu a} \sim&\ \frac{16\pi}{m_{\tilde{a}}} \frac{\Lambda^6v_\PS^2 }{v_H^2 f_\PQ^6 }  \\ \notag
                 \sim&\ 10^{52}~\mathrm{years}\times
                                       \left( \frac{1~\mathrm{keV}}{m_{\tilde{a}}} \right)
                                       \left( \frac{\Lambda}{10^{18}~\mathrm{GeV}} \right)^6
                                       \left( \frac{v_\PS}{10^{16}~\mathrm{GeV}} \right)^2
                                       \left( \frac{10^{10}~\mathrm{GeV}}{f_\PQ} \right)^6.
\end{align}
The second decay mode is similar to the neutralino decay
and can be estimated from Eq.~\eqref{eq-tauchiL} with formally replacing $\chi \to \tilde{a}$,
\begin{align}
 \tau_{\tilde{a}\to \nu\ell\ell}
                   \sim&\ \frac{1536\pi^3}{g_{\nu Z}^4}  \frac{\Lambda^6 v_\PS^2 v_H^2}{f_\PQ^8 m_{\tilde{a}}^3} \\ \notag
                     \sim&\  10^{59}~\mathrm{years}\times
                                       \left(\frac{0.1}{g_{\nu Z}} \right)^4
                                       \left( \frac{1~\mathrm{keV}}{m_{\tilde{a}}} \right)^3
                                       \left( \frac{\Lambda}{10^{18}~\mathrm{GeV}} \right)^6
                                       \left( \frac{v_\PS}{10^{16}~\mathrm{GeV}} \right)^2
                                       \left( \frac{10^{10}~\mathrm{GeV}}{f_\PQ} \right)^8.
\end{align}
Here, the neutrino-axino mixing comes from the second term in Eq.~\eqref{eq-wanu}.
These are both much longer the age of universe,
and thus the axino will be a DM particle if its mass is of $\order{\mathrm{keV}}$ and it is the LSP.

Another way to resolve the overproduction problem is that the LSP is unstable due to sizable RPV
and it does not contribute to the DM.
We will show an example with sizable RPV in the next section.

\subsection{RPV model: low-quality LSP}
\label{sec-rpv}

Let us consider the model with the two PQ fields $P$, $\ol{P}$ and $N_{\ol{\Hcal}} = N_{\ol{\Sigma}} = 1$.
The new fields $\ol{\Hcal}$ and $\ol{\Sigma}$ are mandatory for $h\ne 0$,
because there is no such solution for the anomaly condition in the model only with $\ol{P}$.
The anomaly cancellation conditions are given by
\begin{align}
\label{eq-anomRPV}
 r+\ol{r} \equiv&\ 0 &\quad& \mathrm{modulo}~2,   \\
-3h - p-\ol{p} \equiv&\ -6(h+s)  - 2p  \equiv 6s - 2\ol{p}&\quad&\mathrm{modulo}~N,
\label{eq-anomRPVn}
\end{align}
for $\Zr{4}$, $\Zn{N}$, respectively.

In this section, we shall show an example which violates R-parity
such that the LSP is unstable and decays before BBN.
Let us first consider the RPV effects caused by the RPV Yukawa couplings.
Note that the RPV by $Q\Hcal S^c$ can not be sizable because it does not have $\Zn{N}$ charge,
see Eq.~\eqref{eq-RPV} and Table~\ref{tab-matterPS}~\footnote{
Another bi-linear RPV becomes moderately large
if $(P,\ol{P})^3 Q\ol{\Hcal}S^c$ or $w_0 (P,\ol{P}) Q\ol{\Hcal}S^c$ is allowed.
We do not find any advantage in the first case as discussed in later.
The condition to have the latter is the same as that to have the RPV Yukawa couplings.
}.
The sufficiently large RPV operators are induced if either of the following operators are allowed:
\begin{align}
P QS^cQQ^c:&\ (3+r,-2h+p),\quad& P Q^cQ^cQ^cS^c:&\ (3+r,4s+p),
\end{align}
These operators are allowed by $\Zr{4}$ if $r=3$.
The first one is allowed if $p = 2h$ and the second one is allowed if $p=-4s$.
However, $p=-4s$ is not phenomenologically viable
since there also exists an operator $PS^c\ol{\Sigma}Q^c$ which induces too large a mass
for the down quark $d^c$ of $\order{v_\PS f_\PQ/\Lambda}$.
The same conclusion holds for $\ol{P}$,
and thus the sizable RPV is realized when $P$ or $\ol{P}$ has charge $(3,2h)$.
Since the operator $w_0 Q \ol{\Hcal} S^c$ has charge $(3,-2h)$,
$w_0 P Q \ol{\Hcal}S^c$ ($w_0 \ol{P} Q \ol{\Hcal}S^c$) is accompanied
with $PQS^cQQ^c$ ($\ol{P}QS^cQQ^c$).

\begin{table}[t]
\centering
\caption{\label{tab-RPVPQ}
The charges consistent with the sizable RPV and $b$-term.
$\PQ_\Phi$ is the PQ charge of a field $\Phi$ normalized
such that  the minimal charge of $P$ and $\ol{P}$ is unity.
The PQ charge of $Q^c$ is zero for the Majorana neutrino mass and that of $\ol{\Hcal}$ is opposite to $\Hcal$.
}
\small
\begin{tabular}[t]{c|cc|cc|cc|cccc|c}\hline
 & RPV &$b$-term  & $r$ & $\ol{r}$ & $p$ & $\ol{p}$ & $\PQ_P$ & $\PQ_{\ol{P}}$ & $\PQ_{\Hcal}$ & $\PQ_{Q}$  & $N_{\mathrm{DW}}$
\\ \hline\hline
I &  $PQ^2 Q^cS^c$ & $\ol{P}^2 \Hcal^2$& $3$ & $1$ & $2h$ & $-h$ & $-2$ & $1$& $-1$ &$1$ & 2$\abs{N_g+N_\Psi}$
\\
II & $PQ^2 Q^cS^c$ & $\ol{P}P\Hcal^2$  &  $3$ & $3$ & $2h$ & $-4h$ &  $1$ & $-2$& $1/2$ &$-1/2$& $\abs{2N_\Psi - N_g}$
\\ \hline
III & $\ol{P}Q^2 Q^cS^c$ & $P^2 \Hcal^2$& $1$ & $3$ & $-h$ & $2h$ &  $1$ & $-2$& $-1$ &$1$ & $2\abs{N_g+N_\Psi}$
\\
IV & $\ol{P}Q^2 Q^cS^c$ & $\ol{P}P\Hcal^2$& $3$ & $3$ & $-4h$ & $2h$ &  $-2$ & $1$& $1/2$ &$-1/2$ & $\abs{2N_\Psi-N_g}$
\\    \hline
\end{tabular}
\end{table}

In the non-minimal model,
the higgsino masses are always generated by $W \supset w_0 \Hcal \ol{\Hcal}$.
However, the SUSY breaking $b$-term, $V \supset b_h H_u H_d$, is missing.
For the $b$-term,
$(P,\ol{P})^2 \Hcal^2$ should exist in the superpotential.
We need two PQ fields for the sizable RPV and $b$-term,
since $P^2 \Hcal^2$ is forbidden if $P$ has charge $(3,2h)$.

Table~\ref{tab-RPVPQ} shows four cases
which realize both sizable RPV interaction and $b$-term.
The PQ field $P$ induces the RPV in cases (I) and (II), while $\ol{P}$ does in cases (III) and (IV).
The $b$-term is realized by
$\ol{P}^2\Hcal^2$ in the case (I), $P^2 \Hcal^2$ in the case (III)
and $P\ol{P}\Hcal^2$ in the cases (II) and (IV).
 The PQ charges are determined to be consistent
with interactions in $W_\PS$, $W_\PQ$ and the operators for the sizable RPV and $b$-term.
The axion domain wall number, $N_{\mathrm{DW}}$ is listed in the last row.
For $N_g=3$ and $N_\Psi = 1$, $N_{\mathrm{DW}} = 8$ in the cases (I) and (III),
while $N_{\mathrm{DW}} = 1$ in the cases (II) and (IV).
Thus there is no domain-wall in the latter cases.

The $\Zn{N}$ charge should be chosen such that all the unwanted operators are forbidden by $\Zn{N}$
and satisfy the anomaly cancellation condition Eq.~\eqref{eq-anomRPVn}.  Also
$2h+4s\not\equiv 0$ is required so that $p = 2 h \neq -4 s$, as discussed earlier.
In addition, for proton stability,
if $2h+4s \equiv 0$ then both $P Q^cQ^cQ^c S^c$, which induces baryon number violation,
and $PQ^2Q^cS^c$ which is lepton number violating are allowed.
There are various explicit PQ breaking operators discussed in Section~\ref{sec-PQV}.
In our model search, we set reference values of the scales at
\begin{align}
\label{eq-smaple}
\Lambda=10^{18}~\mathrm{GeV}, \ \
v_\PS =  10^{16}~\mathrm{GeV}, \ \
f_\PQ  = 10^{10}~\mathrm{GeV}, \ \
w_0     = 10^{5}~\mathrm{GeV}, \ \
v_H     =  10^{2}~\mathrm{GeV}.
\end{align}
With these values, we require that the PQ breaking at the tree-level,
Eqs.~\eqref{eq-PQVW0} and~\eqref{eq-PQVK0}, are forbidden such that $\Delta \theta \le 10^{-10}$.
We also require that
the quartic PQ breaking combinations of
$(P, \ol{P})^2 (\Hcal,\ol{\Hcal})^2$ and  $(P, \ol{P})^2 (\Sigma, \oSig)^2$
are forbidden to suppress PQ breaking via the 1-loop effects, see Eq.~\eqref{eq-PQVW1}.
For gauge coupling unification,
$4s \not\equiv 0$ modulo $N$ is required to keep the triplets in the sextets massless at $\order{v_\PS}$.
Although these are still necessary conditions for the fully viable model,
we can find solutions of these conditions only for $N=13$ in the case (II) or $N=15$ in all the four cases
when $N\le 16$ and $N_\Psi = 1$~\footnote{
We can find solutions for $N_\Psi = 2$ when
 $N=11,15$, $N=13, 15$, $N=15$ and $N=13, 15$ in the case (I), (II), (III) and (IV), respectively.
We did the same search for the RPV via $(P,\ol{P})^3 Q\ol{\Hcal} S^c$ in a case of $N_\Psi = 1$,
but we find setups consistent with these conditions only for $N\ge 14$.
We will not study these cases.
}.

\begin{table}[t]
\centering
\caption{\label{tab-RPVexample}
The charges under the discrete symmetries $\Zr{4}\times\Zn{15}$ and an accidental $\UPQ$ symmetry
in our example.
}
\small
\begin{tabular}[t]{c|ccc|cccc|cc|cccc|cc}\hline
  & $\Hcal$   & $Q$ & $Q^c$  & $X$ &$S^c$& $\ol{S}^c$ & $\Sigma$ & $\ol{\Sigma}$& $\ol{\Hcal}$ & $\ol{\Psi}$ & $\Psi$ & ${\Psi}^c$& $\ol{\Psi}^c$ &$P$ & $\ol{P}$ \\ \hline\hline
$\Zbbm_{4R}$&${0}$&  ${1}$&${1}$     & $2$       &  0      &  0    & 2 & $2$ & $0$ & $2$& 1 &1 & 2 & 3 & 3 \\
$\Zbbm_{15}$ &$1$&   $11$ &$3$           & $0$    &  $3$     &  $12$ & $9$ & $6$ & $14$ & $6$ & $13$ & $3$ &$10$ & $11$ & $2$ \\
\hline
$\UPQ$&  $1/2$  & $-1/2$ & $0$   & 0  & $0$  & $0$  & $0$ &  $0$& $-1/2$& $3/2$ & $1/2$ &$0$& $-1$ & $-2$ & $1$   \\
\hline
\end{tabular}
\end{table}

We shall study a solution, $(h,s) = (1,3)$, with $N=15$ in the case (IV)~\footnote{
For $N=13$ in the case (II),
$K_{\cancel{\PQ}}\supset P\ol{P}^\dag \oSig^2$ is always allowed in the K\"{a}hler potential
after imposing the anomaly condition.
This induces $\Delta \theta \sim 10^{-10}$ which is marginal for the quality problem.
}.
The charges of the fields are shown in Table~\ref{tab-RPVexample}.
A complete list of the possibly dangerous operators and their sizes are shown in Appendix~\ref{sec-oplist}.
We discuss operators important for phenomenology in the main text.
The charges of vector-like fields are chosen such that the Yukawa interactions,
\begin{align}
 W_{\mathrm{decay}} = \Psi \ol{\Hcal} Q^c + Q \Hcal \Psi^c,
\end{align}
are allowed by the symmetry so that the vector-like fields decay quickly.
The model with $W_{\mathrm{decay}}\supset \Psi \Hcal Q^c$ instead of $\Psi \ol{\Hcal} Q^c$
also allows an exotic mass term $\ol{\Psi} \Hcal \ol{S}^c:(2,0)$ which gives too large a mass term
for $H_u \Psi_\ell$, where $\Psi_\ell$ is the leptonic component of the vector-like field $\Psi$.
The vector-like triplets from the sextets will decay via $W\supset {w_0}Q^c \Sigma Q^c$
which is allowed by the symmetry independent of the charges $(h,s)$.
The lifetime of the anti-triplet $\ol{\sigma}$ by this interaction is estimated as
\begin{align}
 \tau_{\sigma} \sim  \frac{16\pi  \Lambda^2}{w_0^3}
                \sim  0.01~\mathrm{s} \times \left(\frac{\Lambda}{10^{18}~\mathrm{GeV}}  \right)^2
                                                                 \left(\frac{10^5~\mathrm{GeV}}{w_0}\right)^3,
\end{align}
where the vector-like triplet mass is set at $w_0$.
The triplet $\sigma$ mixes with $\ol{\sigma}$ by the mass term of $\order{w_0}$.
Thus the triplets will decay before BBN if the SUSY scale is as high as $100$~TeV,
and condition (4) is satisfied.

There are four Higgs doublets at the SUSY breaking scale whose mass terms are given by
$W\supset f_\PQ^2 \Hcal^2/\Lambda + w_0 \Hcal \ol{\Hcal}$.
The $b$-term, $V_h \supset b_h H_u H_d$, will be generated after SUSY breaking from the first term.
Therefore, both $\mu$- and $b$-problems are solved in this model
and the condition (3) is satisfied.
Note that we invoke a fine-tuning of $\order{w^2_0/v^2_H} \sim \order{10^6}$ as  usual
in high-scale SUSY breaking scenarios.
More details of the Higgs potential with $\ol{\Hcal}$ is discussed in Appendix~\ref{sec-hhbar}.

The accidental $\UPQ$ charges are shown in the last row of Table~\ref{tab-RPVexample}.
Since the mixed anomaly with $SU(3)_C$ is non-vanishing,
the strong CP problem is solved if the $\UPQ$ symmetry is a sufficiently precise symmetry.
The explicit PQ breaking superpotential is given by
\begin{align}
\label{eq-PQbreakRPV}
W_{\cancel{\mathrm{PQ}}} = \frac{w_0}{\Lambda^{4}} \Hcal^2P^4+\frac{w_0}{\Lambda^{10}}P^9 \ol{P}^3+\cdots.
\end{align}
The full list of PQ breaking operators in the superpotential and K\"{a}hler potential
are shown in Table~\ref{tab-PQVexRPV} of Appendix~\ref{sec-oplist}.
We see that the first term in Eq.~\eqref{eq-PQbreakRPV}
and $K_{\cancel{\PQ}} \supset \Hcal^2P^4$ give $\Delta \theta \sim 10^{-14}$ which are sufficiently small.
Therefore, the PQ symmetry is high quality and the condition (2) is satisfied.

The leading RPV operators and linear terms in the vector-like fields are given by
\begin{align}
\label{eq-leadingRPV}
\Delta W\supset &\ \frac{\ol{P}}{\Lambda^2} Q Q  S^c Q^c
       + \frac{w_0 \ol{P}}{\Lambda^2} Q \ol{\Hcal} S^c
       +  \frac{P\ol{P}^2}{\Lambda^2} \left(
     \ol{S}^c Q^c +
         \frac{1}{\Lambda} Q \Hcal  S^c
       + \frac{1}{\Lambda}  \Psi \ol{\Hcal}  S^c  \right) \\ \notag
  &\    + \frac{w_0}{\Lambda}\left( Q^c \Sigma Q^c + Q\ol{\Sigma}\Psi     \right)
         + \cdots,
\end{align}
where $Q^c$ includes $\Psi^c$ as the fourth family.
The operators in the first line induce RPV effects without baryon number violation.
The first two terms and the first term in the parenthesis
are the dominant source for the lepton number violation, and the others are sub-dominant.
The first term in the parenthesis induces the bi-linear RPV operator $H_u \ell$
after integrating out the right-handed neutrino $\nu^c$.
Its mass parameter is $\order{f_\PQ^3/v_\PS \Lambda}$,
while that for $Q\ol{\Hcal}$ from the second term is $\order{w_0 f_\PQ v_\PS/\Lambda^2}$.
The bi-linear RPV from the second term in the parenthesis is smaller than these contributions.
The last term would induce the bi-linear RPV by the mixing of $Q$  and $\Psi$,
but it is extremely small, unlike the mixing of $Q^c$ and $\Psi^c$,
because the mass parameter for $Q\ol{\Psi}$ is $\order{10^{-22}~\mathrm{GeV}}$.
Note that all of these terms conserve the PQ symmetry,
so that these interactions do not explicitly depend on the axion superfield.
The operators relevant to masses, RPV and proton decay are listed in Table~\ref{tab-RPVexRPV}.
We see that the other mass terms are at most $\order{10^{-13}~\mathrm{GeV}}$ and are negligible.
Therefore, the MSSM particles, $\ol{\Hcal}$ and $(\sigma, \ol{\sigma})$ are lighter than
the SUSY breaking scale $\order{w_0}$, while all the vector-like fields, $\Psi, ~\ol{\Psi}$, have $\order{f_\PQ}$ masses.
The gauge coupling unification holds as discussed in the previous section and the  condition (5) is satisfied.

Except for the operator $Q^c \ol{\Psi}^c$,
the mass terms which contribute to $\mu_T$ and $\mu_{\ol{T}}$ in Eq.~\eqref{eq-VLtriplet}
are smaller than $10^{-21}~\mathrm{GeV}$, see Table~\ref{tab-RPVexRPV}.
These are too small to affect the proton stability.
The mixing via $Q^c \ol{\Psi}^c$ is sizable,
but this can be rotated away by redefining $Q^c$ and $\Psi^c$
without introducing new effects because $\Psi^c$ has the same charges as $Q^c$.
As is explicitly shown in Table~\ref{tab-RPVexRPV},
the coupling constants for the operators linear in the vector-like fields, defined in Eq.~\eqref{eq-VLtriplet},
are $\la_T \lesssim 10^{-18}$ and $\ka_T \lesssim 10^{-31}$ GeV$^{-1}$,
except for those in Eq.~\eqref{eq-leadingRPV}.
Hence, only the operators in Eq.~\eqref{eq-leadingRPV} could induce fast proton decay.
The vector-like pair, $(\Psi,\ol{\Psi})$, is integrated out at $\order{f_\PQ}$
before integrating out $(\sigma, \ol{\sigma})$ whose mass is $\order{w_0}$.
Since $\ol{\Psi}$ and $\ol{\Psi}^c$ are absent in Eq.~\eqref{eq-leadingRPV},
there will be no sizable baryon number violation in the superpotential.
In addition, there cannot be a sizable dimension-6 operator in the K\"{a}hler potential,
since the Yukawa coupling involving the vector-like triplets is at most $\order{10^{-13}}$
from the first term in the second line of Eq.~\eqref{eq-leadingRPV}.
Thus,  R-parity is broken by the lepton number violating operator,
while baryon number is still a precise symmetry such that the proton is stable.

If the neutralino is the LSP, the lifetime of the LSP is as short as $10^{-10}~\mathrm{s}$,
see Eqs.~\eqref{eq-tauchiL} and~\eqref{eq-nhbl},
due to the higgsino-neutrino mixing by the second term in Eq.~\eqref{eq-leadingRPV}.
Thus the neutralino LSP is unstable and will decay before BBN.

The axino LSP may be sufficiently long-lived even with RPV.
The axion superfield $A$ is defined as
\begin{align}
P \sim f_\PQ e^{-2 A/f_\PQ},\quad \ol{P} \sim f_\PQ e^{A/f_\PQ}.
\end{align}
In the RPV model, the axino will mix with higgsinos in the K\"{a}hler potential~\cite{Chun:2006ss},
\begin{align}
  K \supset e^{(A+A^\dag)/2f_\PQ} \Hcal^\dag \Hcal
                    + e^{-(A+A^\dag)/2f_\PQ} \ol{\Hcal}^\dag \ol{\Hcal}  \supset
                            \frac{v_H}{ f_\PQ}   \left( H_d^\dag A + \ol{H}_u ^\dag A\right),
\end{align}
where $\pm 1/2$ in the exponents are the PQ charges of the Higgs bi-doublets.
Note that this mixing with the MSSM fields is absent in the minimal model,
since those are neutral under the PQ symmetry.
Together with the bilinear RPV term, the axino-neutrino mixing arises,
so that the axino will decay via $\tilde{a} \to \nu\ell\ell^\dag$.
The lifetime is estimated as
\begin{align}
 \tau_{\tilde{a}} \sim&\
   \frac{1536\pi^3}{g_{\nu Z}^4}  \frac{v_H^2 v_\PS^2 \Lambda^2}{ f_\PQ^4 m^3_{\tilde{a}}}
             \min\left(1,~\frac{f_\PQ^4 \Lambda^2}{w_0^2 v_\PS^4}\right)  \\ \notag
              \sim&\ 10^{27}~\mathrm{years}
                 \times   \left(\frac{0.1}{g_{\nu Z}}  \right)^4
                 \left(\frac{1~\mathrm{keV}}{m_{\tilde{a}}}\right)^3
                 \left( \frac{\Lambda}{10^{18}~\mathrm{GeV}} \right)^2
                \left( \frac{v_\PS}{10^{16}~\mathrm{GeV}} \right)^2
                \left( \frac{10^{10}~\mathrm{GeV}}{f_\PQ}\right)^4
                ,
\end{align}
where the value of $\min$,
which depends on whether the second or third term in Eq.~\eqref{eq-leadingRPV} dominates
the axino-neutrino mixing,
is taken to be 1 in the second line.
The axino-neutrino mixing induced by the PQ breaking interactions, Eq.~\eqref{eq-PQbreakRPV},
are much more suppressed than that induced  by the RPV, but PQ conserving effects.
In particular, the axion-axino-neutrino Yukawa coupling induced by the PQ breaking interactions
is highly suppressed.
Therefore, the axino lifetime, assuming a mass of order $1~\mathrm{keV}$,
will be much longer than the age of the universe,
although it is much shorter than that in the minimal model.

Altogether, this model has the high-quality axion and the proton is stable,
and thus satisfies all of the conditions (1)-(5).
If the neutralino is the LSP, the overproduction problem is solved because it decays before BBN.
The axion would be the dominant source for the DM.
If the axino is the LSP, the overproduction problem is solved by sufficiently light axino mass
as in the minimal model.
The axino will be a metastable DM particle in addition to the axion DM.

\section{Discussions}
\label{sec-disc}

In this paper,
we proposed supersymmetric Pati-Salam models
with the anomaly-free discrete symmetry $\Zr{4}\times\Zn{N}$.
The anomalous $\UPQ$ symmetry, as well as,   R-parity arise as accidental symmetries
if any of the PQ fields $P$ and $\ol{P}$ have odd R-charge.  We discussed two special models.
In the minimal model, without $\ol{\Hcal}$, $\ol{\Sigma}$, $\ol{P}$,
the anomaly conditions require that the MSSM particles do not carry $\Zn{N}$ charges,
so that the R-parity is respected very accurately and the LSP is sufficiently stable to be the DM.
In the non-minimal cases,
we found an example which violates R-parity such that the neutralino LSP will decay before BBN,
while the accidental $\UPQ$ symmetry is  so accurate that the strong CP problem is solved.
An interesting feature of the RPV case is that the exotic vector-like triplets ($\sigma$, $\ol{\sigma}$)
and bi-doublet $\ol{\Hcal}$ are predicted to have SUSY breaking scale masses.
Since the vector-like triplets may decay through the Yukawa couplings
which are also induced by the SUSY breaking effects,
the SUSY breaking scale is predicted to be larger than $\order{100~\mathrm{TeV}}$.

With the discrete symmetries, there are self-couplings of the PQ field $P$ at very high-order.
It was recently proposed that the baryon asymmetry can be produced through the motion of a PQ field when kicked
by an A-term of the self-coupling $P^n$, so-called, lepto-axiogenesis~\cite{Co:2019wyp,Co:2020jtv}.
Our models may provide concrete examples which can accommodate the lepto-axiogenesis scenario.
In particular, the RPV example will make it easier to explain the relic density of the DM.
A more detailed analysis of lepto-axiogenesis and the phenomenological discussions
about the DM and leptogenesis will be the subject of future work.

\section*{Acknowledgment}
The work of J.K and S.R
is supported in part by the Department of Energy (DOE) under Award No.\ DE-SC0011726.
This work of J.K is supported in part by the Grant-in-Aid for Scientific Research from the
Ministry of Education, Science, Sports and Culture (MEXT), Japan No.\ 18K13534.

\appendix

\section{Higgs sector in the non-minimal model}
\label{sec-hhbar}.

We shall study the Higgs potential with the additional bi-doublet $\ol{\Hcal}$.
We write the bi-doublets by
\begin{align}
 \Hcal =
\begin{pmatrix}
 H_1 \\ H_2
\end{pmatrix},
\quad
\ol{\Hcal} =
\begin{pmatrix}
 H_3 \\ H_4
\end{pmatrix},
\end{align}
where $H_{i}$, $i=1,2,3,4$, are $SU(2)_L$ doublets.
The superpotential is given by
\begin{align}
 W_{\PS} = \frac{1}{2} \mu \Hcal^2  + w_0 \Hcal \ol{\Hcal} \to
 W_H = \mu_{kl} H_k H_l,
\end{align}
where $k=1,3$ and $l=2,4$. In this section, repeated indices are summed over.
With the PS symmetry, $\mu_{12} = \mu$, $\mu_{14} = - \mu_{32} = w_0$ and $\mu_{34} = 0$,
but this relation will not hold after the PS breaking.
The $SU(2)_L$ doublets are contracted by $i\sigma_2$.
We shall study the Higgs potential in the non-minimal model given by
\begin{align}
 V_H = V_{\mathrm{soft}} + V_F + V_D,
\end{align}
with
\begin{align}
 V_{\mathrm{soft}} =&\ \sum_{i=1}^4 m_{H_i}^2 \abs{H_i}^2
                                      +   \left(b_{kl} H_k H_l + h.c. \right), \quad
V_F =  \sum_{k=1,3}  \abs{\mu_{kl} H_l}^2 + \sum_{l=2,4}  \abs{\mu_{kl} H_k}^2   \\
 V_D =&\  \frac{g_1^2}{8} \left(\abs{H_1}^2 - \abs{H_2}^2 + \abs{{H}_3}^2 - \abs{{H}_4}^2 \right)^2
         +  \frac{g_2^2}{2} \left( \sum_{i=1}^4 H_i^* T^a_L H_i \right)^2,
\end{align}
where $T^a_L$, $a=1,2,3$, is generators of $SU(2)_L$.
We first diagonalize the mass terms by redefining the Higgs fields,
\begin{align}
 \begin{pmatrix}
  H_1 \\ {H}_3
 \end{pmatrix}
=:
R_u
 \begin{pmatrix}
  H_u \\ \ol{H}_d
 \end{pmatrix},
\quad
 \begin{pmatrix}
  H_2 \\ {H}_4
 \end{pmatrix}
=:
R_d
 \begin{pmatrix}
  H_d \\ \ol{H}_u
 \end{pmatrix},
\end{align}
where the rotation matrices $R_u$, $R_d$ diagonalize the Higgs mass squared matrices,
\begin{align}
 R_u^\dag
\begin{pmatrix}
 m_{H_1}^2 + \mu^*_{1l} \mu_{1l} & \mu_{1l}^* \mu_{3l}  \\
 \mu_{3l}^* \mu_{1l} &  m_{H_3}^2 +  \mu^*_{3l} \mu_{3l}
\end{pmatrix}
 R_u
=:&\ \mathrm{diag}\left( \mhu, \mbd \right), \\ \notag
 R_d^\dag
\begin{pmatrix}
 m_{H_2}^2 + \mu^*_{k2} \mu_{k2} & \mu^*_{k2} \mu_{k4} \\
 \mu^*_{k4} \mu_{k2}  &  {m}_{H_4}^2 +\mu^*_{k4} \mu_{k4}
\end{pmatrix}
 R_d
=:&\  \mathrm{diag}\left( \mhd, \mbu \right),
\end{align}
where $k = 1,3$ and $l=2,4$ are summed over.
The D-term potential is invariant under this redefinition, so that it is formally replaced by
$ (H_1, H_2, H_3, H_4) \to (H_u, H_d, \Hbd, \Hbu)$.
The $b$-terms are rotated as
\begin{align}
  \begin{pmatrix}
   H_1 & H_3
  \end{pmatrix}
  \begin{pmatrix}
    b_{12} & b_{14} \\ b_{32} & b_{34}
  \end{pmatrix}
  \begin{pmatrix}
   H_2 \\ H_4
  \end{pmatrix}
=
  \begin{pmatrix}
   H_u & \ol{H}_d
  \end{pmatrix}
  \begin{pmatrix}
   b_h & b_u \\ b_d & \ol{b}_h
  \end{pmatrix}
  \begin{pmatrix}
   H_d \\ \ol{H}_u
  \end{pmatrix},
\end{align}
where
\begin{align}
  \begin{pmatrix}
   b_h & b_u \\ b_d & \ol{b}_h
  \end{pmatrix}
:= R_u^T
  \begin{pmatrix}
    b_{12} & b_{14} \\ b_{32} & b_{34}
  \end{pmatrix}
R_d.
\end{align}
The Higgs potential after the rotation is
\begin{align}
 V_H =&\ m^2_{H_u} \abs{H_u}^2 +  m^2_{H_d} \abs{H_d}^2
                  + \ol{m}^2_{H_d} \abs{\ol{H}_d}^2  + \ol{m}^2_{H_u} \abs{\ol{H}_u}^2   \\ \notag
       &+ \left( b_h H_u H_d + b_u H_u \ol{H}_u + b_d  \ol{H}_d H_d + \ol{b}_h \ol{H}_d \ol{H}_u  + h.c. \right)
                  + V_D. 
\end{align}
The first derivatives of the neutral Higgs potential are given by
\begin{align}
 \frac{\partial V_H}{\partial H_u^{0*}} =&\
   \mhu H_u^0   + \la_H \Omega  H_u^0 - b_h H_d^{0*} - b_u \Hbu^{0*},   \\
 \frac{\partial V_H}{\partial H_d^{0*}} =&\
   \mhd H_d^0   - \la_H \Omega  H_d^0  - b_h H_u^{0*} - b_d \Hbd^{0*},  \\
 \frac{\partial V_H}{\partial \Hbd^{0*}} =&\
   \mbd \Hbd^0 + \la_H \Omega  \Hbd^0   - \ol{b}_h \Hbu^{0*}- b_d H_d^{0*},   \\
 \frac{\partial V_H}{\partial \Hbu^{0*}} =&\
   \mbu \Hbu^0  - \la_H \Omega  \Hbu^0  - \ol{b}_h \Hbd^{0*} - b_u H_u^{0*} ,
\end{align}
where
\begin{align}
 \la_H := \frac{g_1^2+g_2^2}{4},\quad
 \Omega := \abs{H_u^0}^2-\abs{H_d^0}^2 +\abs{\Hbd^0}^2 - \abs{\Hbu^0}^2.
\end{align}
The Higgs fields with superscript $0$ are the neutral component of the Higgs doublets.

We define the VEVs and Higgs scalars in the doublets as
\begin{align}
 H_u =&\
\begin{pmatrix}
0 \\  v_u
\end{pmatrix}
+
\frac{1}{\sqrt{2}}
\begin{pmatrix}
 \sqrt{2} H_u^+ \\ h_u + i a_u
\end{pmatrix},
&\quad
 H_d =&\
\begin{pmatrix}
v_d \\  0
\end{pmatrix}
+
\frac{1}{\sqrt{2}}
\begin{pmatrix}
h_d + i a_d   \\  \sqrt{2} H_d^- \\
\end{pmatrix},  \\ \notag
\ol{H}_d =&\
\begin{pmatrix}
0 \\  \ol{v}_d
\end{pmatrix}
+
\frac{1}{\sqrt{2}}
\begin{pmatrix}
 \sqrt{2} \ol{H}_d^+ \\ \ol{h}_d + i \ol{a}_d
\end{pmatrix},
&\quad
\ol{H}_u =&\
\begin{pmatrix}
\ol{v}_u  \\  0
\end{pmatrix}
+
\frac{1}{\sqrt{2}}
\begin{pmatrix}
\ol{h}_u + i \ol{a}_u   \\  \sqrt{2} \ol{H}_u^- \\
\end{pmatrix}.
\end{align}
Elements  of the CP-even mass matrix are given by
\begin{align}
\left[ \Mcal^2_{S} \right]_{h_uh_u}   =&\ 2\la_H v_u^2  + \left(b_h v_d + b_u \ol{v}_u\right)/v_u, &\quad
\left[ \Mcal^2_{S} \right]_{h_uh_d}   =&\ - 2\la_H v_u v_d    - b_h, \\ \notag
\left[ \Mcal^2_{S} \right]_{h_u\ol{h}_d}   =&\ 2\la_H v_u \ol{v}_d, &\quad
\left[ \Mcal^2_{S} \right]_{h_u\ol{h}_u}   =&\ - 2\la_H v_u \ol{v}_u  -b_u, \\ \notag
\left[ \Mcal^2_{S} \right]_{h_dh_d}   =&\ 2\la_H v_d^2 + \left(b_h v_u + b_d \ol{v}_d\right)/v_d, &\quad
\left[ \Mcal^2_{S} \right]_{h_d\ol{h}_d}   =&\ -2\la_H v_d \ol{v}_d -b_d, \\  \notag
\left[ \Mcal^2_{S} \right]_{h_d\ol{h}_u}   =&\ 2\la_H  v_d \ol{v}_u, &\quad
\left[ \Mcal^2_{S} \right]_{\ol{h}_d\ol{h}_d}   =&\ 2\la_H \ol{v}_d^2 + (b_d v_d + \ol{b}_h \ol{v}_u )/\ol{v}_d, \\  \notag
\left[ \Mcal^2_{S} \right]_{\ol{h}_d\ol{h}_u}   =&\ -2  \la \ol{v}_u \ol{v}_d - \ol{b}_h, &\quad
\left[ \Mcal^2_{S} \right]_{\ol{h}_u\ol{h}_u}   =&\ 2\la_h \ol{v}_u^2 + \left(b_u v_u + \ol{b}_u\ol{v}_d \right)/\ol{v}_u,
\end{align}
and those of the CP-odd mass matrix are given by
\begin{align}
\left[ \Mcal^2_{P} \right]_{a_ua_u}   =&\  \left(b_h v_d + b_u \ol{v}_u\right)/v_u,\quad
\left[ \Mcal^2_{P} \right]_{a_ua_d}   =  b_h, \quad
\left[ \Mcal^2_{P} \right]_{a_u\ol{a}_d}   =  0, \quad
\left[ \Mcal^2_{P} \right]_{a_u\ol{a}_u}   = b_u, \notag \\
\left[ \Mcal^2_{P} \right]_{a_da_d}   =&\  \left( b_h v_u + b_d \ol{v}_d\right)/v_d, \quad
\left[ \Mcal^2_{P} \right]_{a_d \ol{a}_d}   = b_d, \quad
\left[ \Mcal^2_{P} \right]_{a_d \ol{a}_u}   =  0, \\  \notag
\left[ \Mcal^2_{P} \right]_{\ol{a}_d\ol{a}_d}   =&\  \left(b_d v_d + \ol{b}_h \ol{v}_u \right)/\ol{v}_d,\quad
\left[ \Mcal^2_{P} \right]_{\ol{a}_d\ol{a}_u}   =  \ol{b}_h, \\ \notag
\left[ \Mcal^2_{P} \right]_{\ol{a}_u\ol{a}_u}   =&\ \left(b_u v_u + \ol{b}_h\ol{v}_d \right)/\ol{v}_u.
\end{align}
Here, the soft masses are replaced by using the minimization conditions.

Let us consider the realistic EW vacuum, $v_u, v_d \gg \ol{v}_u, \ol{v}_d$.
We define
\begin{align}
v_u := v_h s_\beta,\quad
v_d := v_h c_\beta,\quad
\ol{v}_d := \ol{v}_h \ol{c}_\beta,\quad
\ol{v}_u := \ol{v}_h \ol{s}_\beta, \quad
t_\beta := \frac{s_\beta}{c_\beta}, \quad
\ol{t}_\beta := \frac{\ol{s}_\beta}{\ol{c}_\beta}.
\end{align}
Assuming $\mbd, \mbu \gg v_h^2, \ol{v}_h^2$,
the minimization conditions for $\Hbd$, $\Hbu$ become
\begin{align}
\frac{\ol{v}_h}{v_h}
\sim \frac{b_d  c_\beta \ol{c}_\beta - b_u s_\beta \ol{s}_\beta}{\mbd \ol{c}_\beta^2 - \mbu \ol{s}_\beta^2 },
\quad
\ol{t}_\beta \sim \frac{b_d \ol{b}_h +  b_u \mbd t_\beta}{b_d \mbu +  b_u \ol{b}_h t_\beta}.
\end{align}
Thus, $ \mbu, \mbd \gg b_u, b_d$ is required to be $\ol{v}_h \ll v_h$.
Neglecting $\order{\ol{v}_h^2}$, the minimization conditions for $H_u$, $H_d$ are given by
\begin{align}
  \mhu - \la_H v_h^2 c_{2\beta} - b_h / t_\beta =&\ b_u \frac{\ol{v}_h \ol{s}_\beta}{v_h s_\beta}, \\
  \mhd + \la_H v_h^2 c_{2\beta} - b_h  t_\beta =&\ b_d \frac{\ol{v}_h \ol{c}_\beta}{v_h c_\beta}.
\end{align}
The Higgs VEV $v_h$ and vacuum angle $\beta$ obey
\begin{align}
 \la_H v_h^2 = \frac{\tilde{m}_{H_d}^2 - \tilde{m}_{H_u}^2 t_\beta^2 }{t_\beta^2-1},  \quad
 s_{2\beta} = \frac{2b_h}{\tilde{m}_{H_u}^2 + \tilde{m}_{H_d}^2},
\end{align}
where
\begin{align}
 \tilde{m}_{H_u}^2 := \mhu -  b_u \frac{\ol{v}_h \ol{s}_\beta}{v_h s_\beta} ,\quad
 \tilde{m}_{H_d}^2 := \mhd  - b_d \frac{\ol{v}_h \ol{c}_\beta}{v_h c_\beta}.
\end{align}
Note that $\mhu, \mhd$ contain the SUSY contributions from the $\mu$-parameters.
The SUSY breaking parameters should be fine-tuned to realize
$\la_H v_h^2 \sim m_Z^2 = 91.2~\mathrm{GeV}$.
In the RPV model,  $\mu \sim 100$ GeV and $w_0 \sim 10^{5}$ GeV,
so the mixing in $R_u$, $R_d$ are suppressed by $\sim w_0 \mu/ m_{H_{3,4}}^2$.
Hence, the mass parameters $m_{H_u}^2$, $m_{H_d}^2$ which directly relate to the EW scale,
are approximately given by $m_{H_{1,2}}^2 + \abs{w_0}^2 + \abs{\mu}^2$
even if $m^2_{H_{3,4}} \gg b_{u,d}$ for $\ol{v}_h/v_h \ll 1$.
Therefore, the degree of fine-tuning for the EW symmetry breaking
is about $\order{w_0^2/m_Z^2} \sim \order{10^{6}}$.
We also remark that the relation $m_{H_{1,2}}^2 \ll m_{H_{3,4}}^2$
would be naturally realized as a result of renormalization group running
because only $\Hcal$ has the sizable Yukawa couplings with the MSSM quarks.

Before closing, let us estimate the heavy Higgs boson mass spectrum.
Neglecting $\order{v_h^2, b_{u,d} \ol{v}_h/v_h}$, the CP-even and CP-odd mass matrices are given by
\begin{align}
\Mcal_S^2 \sim \Mcal_P^2 \sim
\begin{pmatrix}
 b_h/t_\beta  & b_h & 0 & b_u \\
 b_h & b_h t_\beta & b_d & 0 \\
 0 & b_d & b_d  v_d/\ol{v}_d  & \ol{b}_h \\
b_u & 0 & \ol{b}_h &  b_u v_u / \ol{v}_u
\end{pmatrix}.
\end{align}
Since $\Mcal^2_P = \Mcal^2_S + \order{v_h^2}$ is rank-3, one eigenvalue is zero.
The zero eigenstate in the CP-even Higgs boson corresponds to the SM Higgs boson
whose mass comes from the $\order{v_h^2}$ correction,
and that in the CP-odd Higgs boson corresponds to the NG boson absorbed by a $Z$-boson.
The other three states have masses of $b_h$, $b_d v_d/\ol{v}_d \sim \mbd$
and $b_u v_u/\ol{v}_u \sim \mbu$.

\section{Details of the example models}
\label{sec-oplist}

In this appendix, we shall show possible operators
which can induce the shift in the $\theta$ angle, masses and proton decay.
Their sizes are calculated with
\begin{align}
\Lambda=10^{18}~\mathrm{GeV}, \ \
v_\PS =  10^{16}~\mathrm{GeV}, \ \
f_\PQ  = 10^{10}~\mathrm{GeV}, \ \
w_0     = 10^{5}~\mathrm{GeV}, \ \
v_H     =  10^{2}~\mathrm{GeV}, \notag
\end{align}
as reference values.

Table~\ref{tab-PQVexMIN} lists
the PQ breaking operators and their effects to $\Delta \theta$ in the minimal model.
The tree-level PQ breaking effects are shown in the columns of $\Ocal = \id$.
The others are operators which can affect to the $\theta$ angle through the 1-loop effects.
 We list  all the possible quadratic operators in the superpotential and K\"{a}hler potential.
The cubic operators linearly depending on $S^c$ or $\ol{S}^c$ become quadratic ones after the PS breaking.
For example, we see
\begin{align}
 W_{\cancel{\PQ}} \supset \frac{P^5}{\Lambda^5} \times \Hcal Q S^c
     \sim \frac{f_\PQ^5}{\Lambda^5} v_{\PS} \ell H_u
\end{align}
induces $\Delta \theta \sim 10^{-36}$ at 1-loop level.
The operators which affects $\Delta \theta > 10^{-20}$ are highlighted.
In the minimal model, only the tree-level PQ breaking in the superpotential induces $\Delta\theta > 10^{-20}$.

Tables~\ref{tab-RPVexMIN} shows the operators relevant to the masses (left) and proton decays (right)
in the minimal model.
Similarly to the PQ breaking at 1-loop level,
we show all the quadratic operators and cubic operators which linearly depend on the PS breaking fields.
The columns are highlighted if the mass term is larger than $10^{-9}$ GeV.
The right table shows operators relevant to RPV and proton decay.
The bilinear RPV operators are included in the left table for mass terms.
We studied the dimension-4 RPV, dimension-5 proton decay operators
and those which depend on the vector-like fields linearly.
A column for an operator $\Ocal$ is highlighted
if an effective coupling $\la_{\Ocal} > 10^{-17}$ for dimension-4 operator
or $\ka_{\Ocal} > 10^{-27}~\mathrm{GeV}^{-1}$ for dimension-5 operator.
If all the operators are not highlighted, which is true in the minimal model,
the model satisfy the sufficient conditions to ensure the proton stability as discussed in Section~\ref{sec-BLV}.
Note that the coupling constants include the VEVs of the PS breaking fields.

Tables~\ref{tab-PQVexRPV} and~\ref{tab-RPVexRPV}
show the same as Tables~~\ref{tab-PQVexMIN} and~\ref{tab-RPVexMIN} in the RPV model.
In this analysis, we do not consider the non-zero VEV of $\ol{\Hcal}^2$
because it has the same charge as $w_0 P\ol{P}$
whose VEV is more than 3 orders of magnitude larger than that of $\vev{\ol{\Hcal}^2} \ll v_h^2$.
We see that the shift of the $\theta$ angle is larger than $10^{-20}$ only for the tree-level PQ breaking effects.
The exotic mass terms,
masses not included in $W_\PS, W_\PQ$ nor RPV superpotential in Eq.~\eqref{eq-leadingRPV},
are less than $10^{-13}$ GeV which are negligible.
The cubic and quartic operators for RPV and proton decay
are suppressed as $\la_{\Ocal} \lesssim 10^{-18}$, $\ka_{\Ocal} < 10^{-31}~\mathrm{GeV}^{-1}$
except the couplings in Eq.~\eqref{eq-leadingRPV}.
Most of the operators satisfy the sufficient conditions for proton stability as discussed in Section~\ref{sec-BLV}
and the other operators shown in Eq.~\eqref{eq-leadingRPV} will not induce too fast proton decay
as discussed in Section~\ref{sec-rpv}.
Therefore the proton will not be destabilized by these couplings.

\begin{table}[ph]
 \centering
\caption{\label{tab-PQVexMIN}
Sizes of $\Delta \theta$ in the minimal model.
The operators which induce $\Delta \theta > 10^{-20}$ are highlighted.
}
\begin{minipage}[t]{0.48\hsize}
\footnotesize
 \begin{tabular}[t]{c|c|c}\hline
operator $\Ocal \in W_{\cancel{\PQ}}$ & coupling & $\log_{10} \Delta \theta  $\\  \hline\hline
\rowcolor{HighOrange}  ${\id}$  &  ${P}^{10}$  &  $-17$  \\ \hline
\rowcolor{LightGray}  ${\mathcal{H}}^{2}$  &  ${P}^{10}$  &  $-112$  \\
\rowcolor{LightGray}  $Q^c\overline{S}^c$  &  ${P}^{5}$  &  $-32$  \\
\rowcolor{LightGray}  $Q^c\overline{\Psi}^c$  &  ${P}^{11}w_0$  &  $-154$  \\
\rowcolor{LightGray}  $\overline{S}^cS^c$  &  ${P}^{10}$  &  $-112$  \\
\rowcolor{LightGray}  $\overline{S}^c\Psi^c$  &  ${P}^{5}$  &  $-32$  \\
\rowcolor{LightGray}  $S^c\overline{\Psi}^c$  &  ${P}^{6}$  &  $-48$  \\
\rowcolor{LightGray}  ${X}^{2}$  &  ${P}^{10}$  &  $-112$  \\
\rowcolor{LightGray}  ${\Sigma}^{2}$  &  ${P}^{10}$  &  $-112$  \\
\rowcolor{LightGray}  $Q\overline{\Psi}$  &  ${P}^{11}w_0$  &  $-154$  \\
\rowcolor{LightGray}  $\Psi\overline{\Psi}$  &  ${P}^{11}w_0$  &  $-154$  \\
\rowcolor{LightGray}  $\Psi^c\overline{\Psi}^c$  &  ${P}^{11}w_0$  &  $-154$  \\
\rowcolor{LightGray}  $\mathcal{H}QS^c$  &  ${P}^{5}$  &  $-36$  \\
\rowcolor{LightGray}  $\mathcal{H}S^c\Psi$  &  ${P}^{5}$  &  $-36$  \\
\rowcolor{LightGray}  $Q^cS^c\Sigma$  &  ${P}^{5}w_0$  &  $-62$  \\
\rowcolor{LightGray}  $S^c\Sigma\Psi^c$  &  ${P}^{5}w_0$  &  $-62$  \\
\rowcolor{LightGray}  $\mathcal{H}\overline{S}^c\overline{\Psi}$  &  ${P}^{6}$  &  $-52$  \\
\rowcolor{LightGray}  $\overline{S}^c\Sigma\overline{\Psi}^c$  &  ${P}^{6}w_0$  &  $-78$  \\
\hline
operator $\Ocal\in K_{\cancel{\PQ}}$& coupling & $\log_{10} \Delta \theta$ \\  \hline\hline
\rowcolor{LightGray}  ${\id}$  &  ${P}^{10}w_0$  &  $-43$  \\ \hline
\rowcolor{LightGray}  ${\mathcal{H}}^{2}$  &  ${P}^{10}w_0$  &  $-164$  \\
\rowcolor{LightGray}  $Q^c{Q^c}^{\dagger}$  &  ${P}^{10}w_0$  &  $-164$  \\
\rowcolor{LightGray}  $Q^c\overline{S}^c$  &  ${P}^{5}w_0$  &  $-84$  \\
\rowcolor{LightGray}  $Q^c\overline{\Psi}^c$  &  ${P}^{11}$  &  $-154$  \\
\rowcolor{LightGray}  $Q^c{\Psi^c}^{\dagger}$  &  ${P}^{10}w_0$  &  $-164$  \\
\rowcolor{LightGray}  ${Q^c}^{\dagger}S^c$  &  ${P}^{5}$  &  $-58$  \\
\rowcolor{LightGray}  ${Q^c}^{\dagger}\Psi^c$  &  ${P}^{10}w_0$  &  $-164$  \\
\rowcolor{LightGray}  ${Q^c}^{\dagger}{\overline{\Psi}^c}^{\dagger}$  &  ${P}^{9}$  &  $-122$  \\
\rowcolor{LightGray}  $Q{Q}^{\dagger}$  &  ${P}^{10}w_0$  &  $-164$  \\
\rowcolor{LightGray}  $Q\overline{\Psi}$  &  ${P}^{11}$  &  $-154$  \\
\hline
\end{tabular}
\end{minipage}
\begin{minipage}[t]{0.48\hsize}
\footnotesize
 \begin{tabular}[t]{c|c|c}\hline
operator $\Ocal\in K_{\cancel{\PQ}}$& coupling & $\log_{10} \Delta \theta$ \\  \hline\hline
\rowcolor{LightGray}  $Q{\Psi}^{\dagger}$  &  ${P}^{10}w_0$  &  $-164$  \\
\rowcolor{LightGray}  ${Q}^{\dagger}\Psi$  &  ${P}^{10}w_0$  &  $-164$  \\
\rowcolor{LightGray}  ${Q}^{\dagger}{\overline{\Psi}}^{\dagger}$  &  ${P}^{9}$  &  $-122$  \\
\rowcolor{LightGray}  $\overline{S}^cS^c$  &  ${P}^{10}w_0$  &  $-164$  \\
\rowcolor{LightGray}  $\overline{S}^c\Psi^c$  &  ${P}^{5}w_0$  &  $-84$  \\
\rowcolor{LightGray}  $\overline{S}^c{\overline{\Psi}^c}^{\dagger}$  &  ${P}^{4}$  &  $-42$  \\
\rowcolor{LightGray}  $S^c\overline{\Psi}^c$  &  ${P}^{6}w_0$  &  $-100$  \\
\rowcolor{LightGray}  $S^c{\Psi^c}^{\dagger}$  &  ${P}^{5}$  &  $-58$  \\
\rowcolor{LightGray}  ${X}^{2}$  &  ${P}^{10}w_0$  &  $-164$  \\
\rowcolor{LightGray}  ${\Sigma}^{2}$  &  ${P}^{10}w_0$  &  $-164$  \\
\rowcolor{LightGray}  $\Psi\overline{\Psi}$  &  ${P}^{11}$  &  $-154$  \\
\rowcolor{LightGray}  $\Psi{\Psi}^{\dagger}$  &  ${P}^{10}w_0$  &  $-164$  \\
\rowcolor{LightGray}  $\overline{\Psi}{\overline{\Psi}}^{\dagger}$  &  ${P}^{10}w_0$  &  $-164$  \\
\rowcolor{LightGray}  ${\overline{\Psi}}^{\dagger}{\Psi}^{\dagger}$  &  ${P}^{9}$  &  $-122$  \\
\rowcolor{LightGray}  $\Psi^c\overline{\Psi}^c$  &  ${P}^{11}$  &  $-154$  \\
\rowcolor{LightGray}  $\Psi^c{\Psi^c}^{\dagger}$  &  ${P}^{10}w_0$  &  $-164$  \\
\rowcolor{LightGray}  $\overline{\Psi}^c{\overline{\Psi}^c}^{\dagger}$  &  ${P}^{10}w_0$  &  $-164$  \\
\rowcolor{LightGray}  ${\overline{\Psi}^c}^{\dagger}{\Psi^c}^{\dagger}$  &  ${P}^{9}$  &  $-122$  \\
\rowcolor{LightGray}  $\mathcal{H}QS^c$  &  ${P}^{5}w_0$  &  $-88$  \\
\rowcolor{LightGray}  $\mathcal{H}S^c\Psi$  &  ${P}^{5}w_0$  &  $-88$  \\
\rowcolor{LightGray}  $\mathcal{H}S^c{\overline{\Psi}}^{\dagger}$  &  ${P}^{4}$  &  $-46$  \\
\rowcolor{LightGray}  $Q^cS^c\Sigma$  &  ${P}^{5}$  &  $-62$  \\
\rowcolor{LightGray}  $S^c\Sigma\Psi^c$  &  ${P}^{5}$  &  $-62$  \\
\rowcolor{LightGray}  $S^c\Sigma{\overline{\Psi}^c}^{\dagger}$  &  ${P}^{4}w_0$  &  $-72$  \\
\rowcolor{LightGray}  $\mathcal{H}{Q}^{\dagger}\overline{S}^c$  &  ${P}^{5}$  &  $-62$  \\
\rowcolor{LightGray}  $\mathcal{H}\overline{S}^c\overline{\Psi}$  &  ${P}^{6}w_0$  &  $-104$  \\
\rowcolor{LightGray}  $\mathcal{H}\overline{S}^c{\Psi}^{\dagger}$  &  ${P}^{5}$  &  $-62$  \\
\rowcolor{LightGray}  ${Q^c}^{\dagger}\overline{S}^c\Sigma$  &  ${P}^{5}w_0$  &  $-88$  \\
\rowcolor{LightGray}  $\overline{S}^c\Sigma\overline{\Psi}^c$  &  ${P}^{6}$  &  $-78$  \\
\rowcolor{LightGray}  $\overline{S}^c\Sigma{\Psi^c}^{\dagger}$  &  ${P}^{5}w_0$  &  $-88$  \\
\hline
\end{tabular}
\end{minipage}
\end{table}

\begin{table}[p]
 \centering
\caption{\label{tab-RPVexMIN}
Sizes of masses (left) and coupling constants for dimension-4 and -5 operators
which can be relevant to proton decay(right) in the minimal model.}
\begin{minipage}[t]{0.48\hsize}
\small
 \begin{tabular}[t]{c|c|c}\hline
operator $\Ocal$ & mass $m_\Ocal$& $\log_{10} m_\Ocal  $\\  \hline\hline
\rowcolor{HighOrange}  ${\mathcal{H}}^{2}$  &  $w_0$  &  $5$  \\
\rowcolor{LightGray}  $Q^c\overline{S}^c$  &  ${P}^{5}$  &  $-22$  \\
\rowcolor{HighOrange}  $Q^c\overline{\Psi}^c$  &  $P$  &  $10$  \\
\rowcolor{HighOrange}  $\overline{S}^cS^c$  &  $w_0$  &  $5$  \\
\rowcolor{LightGray}  $\overline{S}^c\Psi^c$  &  ${P}^{5}$  &  $-22$  \\
\rowcolor{LightGray}  $S^c\overline{\Psi}^c$  &  ${P}^{6}$  &  $-30$  \\
\rowcolor{HighOrange}  ${X}^{2}$  &  $w_0$  &  $5$  \\
\rowcolor{HighOrange}  ${\Sigma}^{2}$  &  $w_0$  &  $5$  \\
\rowcolor{HighOrange}  $Q\overline{\Psi}$  &  $P$  &  $10$  \\
\rowcolor{HighOrange}  $\Psi\overline{\Psi}$  &  $P$  &  $10$  \\
\rowcolor{HighOrange}  $\Psi^c\overline{\Psi}^c$  &  $P$  &  $10$  \\  \hline
\rowcolor{HighOrange}  $\overline{S}^cS^cX$  &  $1$  &  $16$  \\
\rowcolor{HighOrange}  $S^cS^c\Sigma$  &  $1$  &  $16$  \\
\rowcolor{HighOrange}  $\overline{S}^c\overline{S}^c\Sigma$  &  $1$  &  $16$  \\
\rowcolor{LightGray}  $\mathcal{H}QS^c$  &  ${P}^{5}$  &  $-24$  \\
\rowcolor{LightGray}  $\mathcal{H}S^c\Psi$  &  ${P}^{5}$  &  $-24$  \\
\rowcolor{LightGray}  $Q^cS^c\Sigma$  &  ${P}^{5}w_0$  &  $-37$  \\
\rowcolor{LightGray}  $S^c\Sigma\Psi^c$  &  ${P}^{5}w_0$  &  $-37$  \\
\rowcolor{LightGray}  $\mathcal{H}\overline{S}^c\overline{\Psi}$  &  ${P}^{6}$  &  $-32$  \\
\rowcolor{LightGray}  $\overline{S}^c\Sigma\overline{\Psi}^c$  &  ${P}^{6}w_0$  &  $-45$  \\
\hline
 \end{tabular}
\end{minipage}
\begin{minipage}[t]{0.48\hsize}
\small
\begin{tabular}[t]{c|c|c}\hline
operator $\Ocal$& $\la_\Ocal$ or $\ka_\Ocal$ & $\log_{10} \la_{\Ocal}, \ka_{\Ocal}$ \\  \hline\hline
\rowcolor{LightGray}  ${Q}^{2}Q^cS^c$  &  ${P}^{5}w_0$  &  $-55$  \\
\rowcolor{LightGray}  ${Q^c}^{3}S^c$  &  ${P}^{5}w_0$  &  $-55$  \\
\rowcolor{LightGray}  ${Q}^{4}$  &  $w_0$  &  $-31$  \\
\rowcolor{LightGray}  ${Q^c}^{4}$  &  $w_0$  &  $-31$  \\  \hline
\rowcolor{LightGray}  ${Q}^{3}\Psi$  &  $w_0$  &  $-31$  \\
\rowcolor{LightGray}  $Q{Q^c}^{2}\Psi$  &  $w_0$  &  $-31$  \\
\rowcolor{LightGray}  $QQ^cS^c\Psi$  &  ${P}^{5}w_0$  &  $-55$  \\
\rowcolor{LightGray}  $QQ^c\overline{S}^c\overline{\Psi}$  &  ${P}^{6}w_0$  &  $-63$  \\
\rowcolor{LightGray}  ${Q}^{2}Q^c\Psi^c$  &  $w_0$  &  $-31$  \\
\rowcolor{LightGray}  ${Q}^{2}S^c\Psi^c$  &  ${P}^{5}w_0$  &  $-55$  \\
\rowcolor{LightGray}  ${Q^c}^{3}\Psi^c$  &  $w_0$  &  $-31$  \\
\rowcolor{LightGray}  ${Q^c}^{2}S^c\Psi^c$  &  ${P}^{5}w_0$  &  $-55$  \\
\rowcolor{LightGray}  ${Q}^{2}\overline{S}^c\overline{\Psi}^c$  &  ${P}^{6}w_0$  &  $-63$  \\
\rowcolor{LightGray}  ${Q^c}^{2}\overline{S}^c\overline{\Psi}^c$  &  ${P}^{6}w_0$  &  $-63$  \\
\hline
\end{tabular}
\end{minipage}
\end{table}

\begin{table}[p]
 \centering
\caption{\label{tab-PQVexRPV}
Sizes of $\Delta \theta$ in the RPV model.
The operators which induce $\Delta \theta > 10^{-20}$ are highlighted.
}
\begin{minipage}[t]{0.48\hsize}
\scriptsize
 \begin{tabular}[t]{c|c|c}\hline
operator $\Ocal \in W_{\cancel{\PQ}}$ & coupling & $\log_{10} \Delta \theta  $\\  \hline\hline
\rowcolor{HighOrange}  ${\id}$  &  ${\mathcal{H}}^{2}{P}^{4}w_0$  &  $-14$  \\ \hline
\rowcolor{LightGray}  ${\mathcal{H}}^{2}$  &  ${P}^{8}w_0$  &  $-106$  \\
\rowcolor{LightGray}  $\overline{\mathcal{H}}\mathcal{H}$  &  ${\mathcal{H}}^{2}{P}^{4}w_0$  &  $-106$  \\
\rowcolor{LightGray}  ${\overline{\mathcal{H}}}^{2}$  &  ${P}^{8}{\overline{P}}^{2}$  &  $-112$  \\
\rowcolor{LightGray}  $Q^c\overline{S}^c$  &  ${P}^{8}\overline{P}w_0$  &  $-122$  \\
\rowcolor{LightGray}  $Q^c\overline{\Psi}^c$  &  ${P}^{7}w_0$  &  $-90$  \\
\rowcolor{LightGray}  $\overline{S}^cS^c$  &  ${\mathcal{H}}^{2}{P}^{4}w_0$  &  $-106$  \\
\rowcolor{LightGray}  $\overline{S}^c\Psi^c$  &  ${P}^{8}\overline{P}w_0$  &  $-122$  \\
\rowcolor{LightGray}  $S^c\overline{\Psi}^c$  &  ${P}^{8}{\overline{P}}^{2}w_0$  &  $-138$  \\
\rowcolor{LightGray}  ${X}^{2}$  &  ${\mathcal{H}}^{2}{P}^{4}w_0$  &  $-106$  \\
\rowcolor{LightGray}  ${\Sigma}^{2}$  &  ${\overline{P}}^{6}$  &  $-48$  \\
\rowcolor{LightGray}  $\Sigma\overline{\Sigma}$  &  ${\mathcal{H}}^{2}{P}^{4}w_0$  &  $-106$  \\
\rowcolor{LightGray}  ${\overline{\Sigma}}^{2}$  &  ${P}^{4}{\overline{P}}^{2}$  &  $-48$  \\
\rowcolor{LightGray}  $Q\overline{\Psi}$  &  ${P}^{9}{\overline{P}}^{2}w_0$  &  $-154$  \\
\rowcolor{LightGray}  $\Psi\overline{\Psi}$  &  ${\mathcal{H}}^{2}{\overline{P}}^{5}$  &  $-96$  \\
\rowcolor{LightGray}  $\Psi^c\overline{\Psi}^c$  &  ${P}^{7}w_0$  &  $-90$  \\
\rowcolor{LightGray}  $\mathcal{H}QS^c$  &  ${P}^{8}\overline{P}w_0$  &  $-126$  \\
\rowcolor{LightGray}  $\mathcal{H}S^c\Psi$  &  ${P}^{9}{\overline{P}}^{2}$  &  $-132$  \\
\rowcolor{LightGray}  $\overline{\mathcal{H}}QS^c$  &  ${P}^{7}$  &  $-68$  \\
\rowcolor{LightGray}  $\overline{\mathcal{H}}S^c\Psi$  &  ${P}^{8}\overline{P}w_0$  &  $-126$  \\
\rowcolor{LightGray}  $Q^cS^c\Sigma$  &  ${P}^{8}\overline{P}$  &  $-100$  \\
\rowcolor{LightGray}  $Q^cS^c\overline{\Sigma}$  &  ${P}^{3}w_0$  &  $-30$  \\
\rowcolor{LightGray}  $S^c\Sigma\Psi^c$  &  ${P}^{8}\overline{P}$  &  $-100$  \\
\rowcolor{LightGray}  $S^c\overline{\Sigma}\Psi^c$  &  ${P}^{3}w_0$  &  $-30$  \\
\rowcolor{LightGray}  $\mathcal{H}\overline{S}^c\overline{\Psi}$  &  ${P}^{9}\overline{P}w_0$  &  $-142$  \\
\rowcolor{LightGray}  $\overline{\mathcal{H}}\overline{S}^c\overline{\Psi}$  &  ${P}^{8}$  &  $-84$  \\
\rowcolor{LightGray}  $\overline{S}^c\Sigma\overline{\Psi}^c$  &  ${P}^{4}w_0$  &  $-46$  \\
\rowcolor{LightGray}  $\overline{S}^c\overline{\Sigma}\overline{\Psi}^c$  &  ${P}^{8}{\overline{P}}^{2}$  &  $-116$  \\
\hline
operator $\Ocal\in K_{\cancel{\PQ}}$& coupling & $\log_{10} \Delta \theta$ \\  \hline\hline
\rowcolor{HighOrange}  ${\id}$  &  ${\mathcal{H}}^{2}{P}^{4}$  &  $-14$  \\ \hline
\rowcolor{LightGray}  ${\mathcal{H}}^{2}$  &  ${P}^{8}$  &  $-106$  \\
\rowcolor{LightGray}  $\overline{\mathcal{H}}\mathcal{H}$  &  ${\mathcal{H}}^{2}{P}^{4}$  &  $-106$  \\
\rowcolor{LightGray}  ${\overline{\mathcal{H}}}^{2}$  &  ${P}^{6}{{\overline{P}}^{\dagger}}^{2}$  &  $-106$  \\
\rowcolor{LightGray}  $Q^c{Q^c}^{\dagger}$  &  ${\mathcal{H}}^{2}{P}^{4}$  &  $-106$  \\
\rowcolor{LightGray}  $Q^c\overline{S}^c$  &  ${\mathcal{H}}^{2}{P}^{3}{{\overline{P}}^{\dagger}}^{2}$  &  $-122$  \\
\rowcolor{LightGray}  $Q^c\overline{\Psi}^c$  &  ${P}^{7}$  &  $-90$  \\
\rowcolor{LightGray}  $Q^c{\Psi^c}^{\dagger}$  &  ${\mathcal{H}}^{2}{P}^{4}$  &  $-106$  \\
\rowcolor{LightGray}  ${Q^c}^{\dagger}S^c$  &  ${P}^{6}{{\overline{P}}^{\dagger}}^{3}$  &  $-122$  \\
\rowcolor{LightGray}  ${Q^c}^{\dagger}\Psi^c$  &  ${\mathcal{H}}^{2}{P}^{4}$  &  $-106$  \\
\rowcolor{LightGray}  ${Q^c}^{\dagger}{\overline{\Psi}^c}^{\dagger}$  &  ${P}^{7}{{\overline{P}}^{\dagger}}^{2}$  &  $-122$  \\
\rowcolor{LightGray}  $Q{Q}^{\dagger}$  &  ${\mathcal{H}}^{2}{P}^{4}$  &  $-106$  \\
\rowcolor{LightGray}  $Q\overline{\Psi}$  &  ${\mathcal{H}}^{2}{P}^{4}{\overline{P}}^{\dagger}$  &  $-122$  \\
\rowcolor{LightGray}  $Q{\Psi}^{\dagger}$  &  ${P}^{6}{{\overline{P}}^{\dagger}}^{2}$  &  $-106$  \\
\rowcolor{LightGray}  ${Q}^{\dagger}\Psi$  &  ${P}^{8}$  &  $-106$  \\
\hline
\end{tabular}
\end{minipage}
\begin{minipage}[t]{0.48\hsize}
\scriptsize
 \begin{tabular}[t]{c|c|c}\hline
operator $\Ocal\in K_{\cancel{\PQ}}$& coupling & $\log_{10} \Delta \theta$ \\  \hline\hline
\rowcolor{LightGray}  ${Q}^{\dagger}{\overline{\Psi}}^{\dagger}$  &  ${P}^{7}w_0$  &  $-116$  \\
\rowcolor{LightGray}  $\overline{S}^cS^c$  &  ${\mathcal{H}}^{2}{P}^{4}$  &  $-106$  \\
\rowcolor{LightGray}  $\overline{S}^c\Psi^c$  &  ${\mathcal{H}}^{2}{P}^{3}{{\overline{P}}^{\dagger}}^{2}$  &  $-122$  \\
\rowcolor{LightGray}  $\overline{S}^c{\overline{\Psi}^c}^{\dagger}$  &  ${P}^{8}w_0$  &  $-132$  \\
\rowcolor{LightGray}  $S^c\overline{\Psi}^c$  &  ${\mathcal{H}}^{2}{P}^{3}{\overline{P}}^{\dagger}$  &  $-106$  \\
\rowcolor{LightGray}  $S^c{\Psi^c}^{\dagger}$  &  ${P}^{6}{{\overline{P}}^{\dagger}}^{3}$  &  $-122$  \\
\rowcolor{LightGray}  ${X}^{2}$  &  ${\mathcal{H}}^{2}{P}^{4}$  &  $-106$  \\
\rowcolor{LightGray}  ${\Sigma}^{2}$  &  ${P}^{3}{{\overline{P}}^{\dagger}}^{3}$  &  $-74$  \\
\rowcolor{LightGray}  $\Sigma\overline{\Sigma}$  &  ${\mathcal{H}}^{2}{P}^{4}$  &  $-106$  \\
\rowcolor{LightGray}  ${\overline{\Sigma}}^{2}$  &  ${P}^{2}{{\overline{P}}^{\dagger}}^{2}$  &  $-42$  \\
\rowcolor{LightGray}  $\Psi\overline{\Psi}$  &  ${P}^{8}{\overline{P}}^{\dagger}$  &  $-122$  \\
\rowcolor{LightGray}  $\Psi{\Psi}^{\dagger}$  &  ${\mathcal{H}}^{2}{P}^{4}$  &  $-106$  \\
\rowcolor{LightGray}  $\overline{\Psi}{\overline{\Psi}}^{\dagger}$  &  ${\mathcal{H}}^{2}{P}^{4}$  &  $-106$  \\
\rowcolor{LightGray}  ${\overline{\Psi}}^{\dagger}{\Psi}^{\dagger}$  &  ${P}^{6}{\overline{P}}^{\dagger}$  &  $-90$  \\
\rowcolor{LightGray}  $\Psi^c\overline{\Psi}^c$  &  ${P}^{7}$  &  $-90$  \\
\rowcolor{LightGray}  $\Psi^c{\Psi^c}^{\dagger}$  &  ${\mathcal{H}}^{2}{P}^{4}$  &  $-106$  \\
\rowcolor{LightGray}  $\overline{\Psi}^c{\overline{\Psi}^c}^{\dagger}$  &  ${\mathcal{H}}^{2}{P}^{4}$  &  $-106$  \\
\rowcolor{LightGray}  ${\overline{\Psi}^c}^{\dagger}{\Psi^c}^{\dagger}$  &  ${P}^{7}{{\overline{P}}^{\dagger}}^{2}$  &  $-122$  \\
\rowcolor{LightGray}  $\mathcal{H}QS^c$  &  ${\mathcal{H}}^{2}{P}^{3}{{\overline{P}}^{\dagger}}^{2}$  &  $-126$  \\
\rowcolor{LightGray}  $\mathcal{H}S^c\Psi$  &  ${P}^{7}{{\overline{P}}^{\dagger}}^{2}$  &  $-126$  \\
\rowcolor{LightGray}  $\mathcal{H}S^c{\overline{\Psi}}^{\dagger}$  &  ${\mathcal{H}}^{2}{P}^{3}{\overline{P}}^{\dagger}$  &  $-110$  \\
\rowcolor{LightGray}  $\overline{\mathcal{H}}QS^c$  &  ${P}^{7}w_0$  &  $-120$  \\
\rowcolor{LightGray}  $\overline{\mathcal{H}}S^c\Psi$  &  ${\mathcal{H}}^{2}{P}^{3}{{\overline{P}}^{\dagger}}^{2}$  &  $-126$  \\
\rowcolor{LightGray}  $\overline{\mathcal{H}}S^c{\overline{\Psi}}^{\dagger}$  &  ${P}^{5}{{\overline{P}}^{\dagger}}^{3}$  &  $-110$  \\
\rowcolor{LightGray}  $Q^cS^c\Sigma$  &  ${P}^{6}{{\overline{P}}^{\dagger}}^{3}$  &  $-126$  \\
\rowcolor{LightGray}  $Q^cS^c\overline{\Sigma}$  &  ${P}^{3}$  &  $-30$  \\
\rowcolor{LightGray}  $S^c\Sigma\Psi^c$  &  ${P}^{6}{{\overline{P}}^{\dagger}}^{3}$  &  $-126$  \\
\rowcolor{LightGray}  $S^c\Sigma{\overline{\Psi}^c}^{\dagger}$  &  ${P}^{8}$  &  $-110$  \\
\rowcolor{LightGray}  $S^c\overline{\Sigma}\Psi^c$  &  ${P}^{3}$  &  $-30$  \\
\rowcolor{LightGray}  $S^c\overline{\Sigma}{\overline{\Psi}^c}^{\dagger}$  &  ${\mathcal{H}}^{2}$  &  $-46$  \\
\rowcolor{LightGray}  $\mathcal{H}{Q}^{\dagger}\overline{S}^c$  &  ${\mathcal{H}}^{2}{P}^{4}{\overline{P}}^{\dagger}$  &  $-126$  \\
\rowcolor{LightGray}  $\mathcal{H}\overline{S}^c\overline{\Psi}$  &  ${\mathcal{H}}^{2}{P}^{4}{{\overline{P}}^{\dagger}}^{2}$  &  $-142$  \\
\rowcolor{LightGray}  $\mathcal{H}\overline{S}^c{\Psi}^{\dagger}$  &  ${P}^{6}{{\overline{P}}^{\dagger}}^{3}$  &  $-126$  \\
\rowcolor{LightGray}  $\overline{\mathcal{H}}{Q}^{\dagger}\overline{S}^c$  &  ${P}^{6}{{\overline{P}}^{\dagger}}^{3}$  &  $-126$  \\
\rowcolor{LightGray}  $\overline{\mathcal{H}}\overline{S}^c\overline{\Psi}$  &  ${P}^{8}w_0$  &  $-136$  \\
\rowcolor{LightGray}  $\overline{\mathcal{H}}\overline{S}^c{\Psi}^{\dagger}$  &  ${P}^{7}$  &  $-94$  \\
\rowcolor{LightGray}  ${Q^c}^{\dagger}\overline{S}^c\Sigma$  &  ${\mathcal{H}}^{2}P$  &  $-62$  \\
\rowcolor{LightGray}  ${Q^c}^{\dagger}\overline{S}^c\overline{\Sigma}$  &  ${\mathcal{H}}^{2}{P}^{3}{{\overline{P}}^{\dagger}}^{2}$  &  $-126$  \\
\rowcolor{LightGray}  $\overline{S}^c\Sigma\overline{\Psi}^c$  &  ${P}^{4}$  &  $-46$  \\
\rowcolor{LightGray}  $\overline{S}^c\Sigma{\Psi^c}^{\dagger}$  &  ${\mathcal{H}}^{2}P$  &  $-62$  \\
\rowcolor{LightGray}  $\overline{S}^c\overline{\Sigma}\overline{\Psi}^c$  &  ${P}^{6}{{\overline{P}}^{\dagger}}^{2}$  &  $-110$  \\
\rowcolor{LightGray}  $\overline{S}^c\overline{\Sigma}{\Psi^c}^{\dagger}$  &  ${\mathcal{H}}^{2}{P}^{3}{{\overline{P}}^{\dagger}}^{2}$  &  $-126$  \\
\hline
\end{tabular}
\end{minipage}
\end{table}

\begin{table}[p]
 \centering
\caption{\label{tab-RPVexRPV}
Sizes of masses (left) and coupling constants for dimension-4 and -5 operators
which can be relevant to proton decay (right) in the RPV model.
}
\begin{minipage}[t]{0.48\hsize}
\small
 \begin{tabular}[t]{c|c|c}\hline
operator $\Ocal$ & mass $m_\Ocal$& $\log_{10} m_\Ocal  $\\  \hline\hline
\rowcolor{HighOrange}  ${\mathcal{H}}^{2}$  &  $P\overline{P}$  &  $2$  \\
\rowcolor{HighOrange}  $\overline{\mathcal{H}}\mathcal{H}$  &  $w_0$  &  $5$  \\
\rowcolor{LightGray}  ${\overline{\mathcal{H}}}^{2}$  &  $P{\overline{P}}^{3}w_0$  &  $-27$  \\
\rowcolor{HighOrange}  $Q^c\overline{S}^c$  &  $P{\overline{P}}^{2}$  &  $-6$  \\
\rowcolor{HighOrange}  $Q^c\overline{\Psi}^c$  &  $\overline{P}$  &  $10$  \\
\rowcolor{HighOrange}  $\overline{S}^cS^c$  &  $w_0$  &  $5$  \\
\rowcolor{HighOrange}  $\overline{S}^c\Psi^c$  &  $P{\overline{P}}^{2}$  &  $-6$  \\
\rowcolor{LightGray}  $S^c\overline{\Psi}^c$  &  $P{\overline{P}}^{3}$  &  $-14$  \\
\rowcolor{HighOrange}  ${X}^{2}$  &  $w_0$  &  $5$  \\
\rowcolor{LightGray}  ${\Sigma}^{2}$  &  ${\overline{P}}^{6}$  &  $-30$  \\
\rowcolor{HighOrange}  $\Sigma\overline{\Sigma}$  &  $w_0$  &  $5$  \\
\rowcolor{LightGray}  ${\overline{\Sigma}}^{2}$  &  ${P}^{4}{\overline{P}}^{2}$  &  $-30$  \\
\rowcolor{LightGray}  $Q\overline{\Psi}$  &  ${P}^{2}{\overline{P}}^{3}$  &  $-22$  \\
\rowcolor{HighOrange}  $\Psi\overline{\Psi}$  &  $P$  &  $10$  \\
\rowcolor{HighOrange}  $\Psi^c\overline{\Psi}^c$  &  $\overline{P}$  &  $10$  \\ \hline
\rowcolor{HighOrange}  $\overline{S}^cS^cX$  &  $1$  &  $16$  \\
\rowcolor{HighOrange}  ${S}^cS^c\Sigma$  &  $1$  &  $16$  \\
\rowcolor{HighOrange}  $\overline{S}^c\overline{S}^c \ol{\Sigma}$  &  $1$  &  $16$  \\
\rowcolor{HighOrange}  $\mathcal{H}QS^c$  &  $P{\overline{P}}^{2}$  &  $-8$  \\
\rowcolor{LightGray}  $\mathcal{H}S^c\Psi$  &  ${P}^{2}{\overline{P}}^{3}w_0$  &  $-37$  \\
\rowcolor{HighOrange}  $\overline{\mathcal{H}}QS^c$  &  $\overline{P}w_0$  &  $-5$  \\
\rowcolor{HighOrange}  $\overline{\mathcal{H}}S^c\Psi$  &  $P{\overline{P}}^{2}$  &  $-8$  \\
\rowcolor{LightGray}  $Q^cS^c\Sigma$  &  $P{\overline{P}}^{2}w_0$  &  $-21$  \\
\rowcolor{LightGray}  $Q^cS^c\overline{\Sigma}$  &  ${P}^{3}w_0$  &  $-21$  \\
\rowcolor{LightGray}  $S^c\Sigma\Psi^c$  &  $P{\overline{P}}^{2}w_0$  &  $-21$  \\
\rowcolor{LightGray}  $S^c\overline{\Sigma}\Psi^c$  &  ${P}^{3}w_0$  &  $-21$  \\
\rowcolor{LightGray}  $\mathcal{H}\overline{S}^c\overline{\Psi}$  &  ${P}^{2}{\overline{P}}^{2}$  &  $-16$  \\
\rowcolor{LightGray}  $\overline{\mathcal{H}}\overline{S}^c\overline{\Psi}$  &  $P\overline{P}w_0$  &  $-13$  \\
\rowcolor{LightGray}  $\overline{S}^c\Sigma\overline{\Psi}^c$  &  ${P}^{4}w_0$  &  $-29$  \\
\rowcolor{LightGray}  $\overline{S}^c\overline{\Sigma}\overline{\Psi}^c$  &  $P{\overline{P}}^{3}w_0$  &  $-29$  \\
\hline
 \end{tabular}
\end{minipage}
\begin{minipage}[t]{0.48\hsize}
\small
\begin{tabular}[t]{c|c|c}\hline
operator $\Ocal$& $\la_\Ocal$ or $\ka_\Ocal$ & $\log_{10} \la_{\Ocal}, \ka_{\Ocal}$ \\  \hline\hline
\rowcolor{HighOrange}  ${Q}^{2}Q^cS^c$  &  $\overline{P}$  &  $-10$  \\
\rowcolor{LightGray}  ${Q^c}^{3}S^c$  &  ${P}^{3}w_0$  &  $-39$  \\
\rowcolor{LightGray}  ${Q}^{4}$  &  ${\mathcal{H}}^{2}w_0$  &  $-63$  \\
\rowcolor{LightGray}  ${Q^c}^{4}$  &  ${P}^{4}{\overline{P}}^{2}$  &  $-66$  \\ \hline
\rowcolor{LightGray}  $Q\Sigma\Psi$  &  ${\overline{P}}^{6}$  &  $-48$  \\
\rowcolor{HighOrange}  $Q\overline{\Sigma}\Psi$  &  $w_0$  &  $-13$  \\
\rowcolor{LightGray}  ${Q}^{3}\Psi$  &  ${P}^{4}w_0$  &  $-63$  \\
\rowcolor{LightGray}  $Q{Q^c}^{2}\Psi$  &  $w_0$  &  $-31$  \\
\rowcolor{LightGray}  $QQ^cS^c\Psi$  &  $P{\overline{P}}^{2}w_0$  &  $-39$  \\
\rowcolor{LightGray}  $QQ^c\overline{S}^c\Sigma\Psi$  &  ${\mathcal{H}}^{2}Pw_0$  &  $-73$  \\ \hline
\rowcolor{LightGray}  $QQ^c\overline{S}^c\overline{\Psi}$  &  $P\overline{P}$  &  $-18$  \\
\rowcolor{LightGray}  $QQ^cS^c\overline{\Sigma}\overline{\Psi}$  &  ${\mathcal{H}}^{2}$  &  $-52$  \\ \hline
\rowcolor{HighOrange}  $Q^c\Sigma\Psi^c$  &  $w_0$  &  $-13$  \\
\rowcolor{LightGray}  $Q^c\overline{\Sigma}\Psi^c$  &  ${P}^{4}{\overline{P}}^{2}$  &  $-48$  \\
\rowcolor{LightGray}  ${Q}^{2}Q^c\Psi^c$  &  $P{\overline{P}}^{3}w_0$  &  $-63$  \\
\rowcolor{HighOrange}  ${Q}^{2}S^c\Psi^c$  &  $\overline{P}$  &  $-10$  \\
\rowcolor{LightGray}  ${Q^c}^{3}\Psi^c$  &  ${P}^{4}{\overline{P}}^{2}$  &  $-66$  \\
\rowcolor{LightGray}  ${Q^c}^{2}S^c\Psi^c$  &  ${P}^{3}w_0$  &  $-39$  \\
\rowcolor{LightGray}  ${Q}^{2}\overline{S}^c\Sigma\Psi^c$  &  ${\overline{P}}^{7}$  &  $-76$  \\
\rowcolor{LightGray}  ${Q^c}^{2}\overline{S}^c\Sigma\Psi^c$  &  $P{\overline{P}}^{2}$  &  $-44$  \\ \hline
\rowcolor{LightGray}  ${Q}^{2}\overline{S}^c\overline{\Psi}^c$  &  ${\mathcal{H}}^{2}w_0$  &  $-47$  \\
\rowcolor{LightGray}  ${Q^c}^{2}\overline{S}^c\overline{\Psi}^c$  &  $P{\overline{P}}^{3}w_0$  &  $-47$  \\
\rowcolor{LightGray}  ${Q}^{2}S^c\overline{\Sigma}\overline{\Psi}^c$  &  ${\overline{P}}^{2}w_0$  &  $-49$  \\
\rowcolor{LightGray}  ${Q^c}^{2}S^c\overline{\Sigma}\overline{\Psi}^c$  &  ${P}^{3}\overline{P}$  &  $-52$  \\ \hline
\rowcolor{LightGray}  ${Q}^{2}\Sigma$  &  ${P}^{4}w_0$  &  $-45$  \\
\rowcolor{HighOrange}  ${Q^c}^{2}\Sigma$  &  $w_0$  &  $-13$  \\
\rowcolor{LightGray}  ${Q}^{2}Q^c\overline{S}^c\Sigma$  &  ${\overline{P}}^{7}$  &  $-76$  \\
\rowcolor{LightGray}  ${Q^c}^{3}\overline{S}^c\Sigma$  &  $P{\overline{P}}^{2}$  &  $-44$  \\ \hline
\rowcolor{LightGray}  ${Q}^{2}\overline{\Sigma}$  &  $P{\overline{P}}^{3}w_0$  &  $-45$  \\
\rowcolor{LightGray}  ${Q^c}^{2}\overline{\Sigma}$  &  ${P}^{4}{\overline{P}}^{2}$  &  $-48$  \\
\rowcolor{LightGray}  ${Q}^{2}Q^c\overline{S}^c\overline{\Sigma}$  &  $\overline{P}w_0$  &  $-41$  \\
\rowcolor{LightGray}  ${Q^c}^{3}\overline{S}^c\overline{\Sigma}$  &  ${P}^{3}$  &  $-44$  \\
\hline
\end{tabular}
\end{minipage}
\end{table}

\clearpage
{\small
\bibliographystyle{JHEP}
\bibliography{PSleptogenesis}
}

\end{document}